\documentclass[%
 superscriptaddress,
 reprint,
 showpacs,preprintnumbers,
 nofootinbib,
 amsmath,amssymb,
 aps,
 prd,
 floatfix,
]{revtex4-1}
\usepackage{graphicx}
\usepackage{xcolor}
\usepackage{hyperref}
\hypersetup{
    colorlinks=true,
    linkcolor=blue,
    filecolor=blue,      
    urlcolor=blue,
    pdftitle={ND280 Upgrade Sensitivity},
    }

\usepackage{amssymb,amsmath,amstext,amsthm,amsfonts}
\usepackage[utf8]{inputenc}
\usepackage{float}
\usepackage{url}
\usepackage{soul}

\RequirePackage{xspace}

\newcommand{\sfgd}{Super-FGD\xspace}
\newcommand{\hatpc}{HA-TPC\xspace}

\def\nub        {\ensuremath{\overline{\nu}}\xspace}

\usepackage{lineno}

\definecolor{LightCyan}{rgb}{0.88,1,1}
\usepackage{xcolor,colortbl}

\begin{document}

	\title{Sensitivity of the Upgraded T2K Near Detector to constrain neutrino and anti-neutrino interactions with no mesons in the final state by exploiting nucleon-lepton correlations}

%

	\author{S.~Dolan}
	\email[Contact e-mail: ]{Stephen.Joseph.Dolan@cern.ch}
	\affiliation{European Organization for Nuclear Research (CERN), 1211 Geneva 23, Switzerland}
	
	\author{V.~Q.~Nguyen}
	\email[Contact e-mail: ]{quocviet.nguyen@lpnhe.in2p3.fr}
	\affiliation{LPNHE, Sorbonne Universit\'e, Universit\'e de Paris, CNRS/IN2P3, Paris, France}

	\author{A.~Blanchet}
	\affiliation{LPNHE, Sorbonne Universit\'e, Universit\'e de Paris, CNRS/IN2P3, Paris, France}

	\author{S.~Bolognesi}
	\affiliation{IRFU, CEA, Universit\'e Paris-Saclay, Gif-sur-Yvette, France}

    \author{M.~Buizza Avanzini}
	\affiliation{Laboratoire Leprince-Ringuet, CNRS, Ecole polytechnique, Institut Polytechnique de Paris, Palaiseau, France}
	
    \author{J.~Chakrani}
	\affiliation{Laboratoire Leprince-Ringuet, CNRS, Ecole polytechnique, Institut Polytechnique de Paris, Palaiseau, France}
	
	\author{A.~Ershova}
	\affiliation{IRFU, CEA, Universit\'e Paris-Saclay, Gif-sur-Yvette, France}

	\author{C.~Giganti}
	\affiliation{LPNHE, Sorbonne Universit\'e, Universit\'e de Paris, CNRS/IN2P3, Paris, France}
	
	\author{Y.~Kudenko}
	\affiliation{Institute for Nuclear Research of the Russian Academy of Sciences, Moscow, Russia}	
	\affiliation{Moscow Institute of Physics and Technology (MIPT), Moscow, Russia}		
	\affiliation{National Research Nuclear University MEPhI, Moscow, Russia}

	\author{M.~Lamoureux}
	\affiliation{INFN Sezione di Padova, I-35131, Padova, Italy}

	\author{A.~Letourneau}
	\affiliation{IRFU, CEA, Universit\'e Paris-Saclay, Gif-sur-Yvette, France}
	
	\author{M.~Martini}
	\affiliation{LPNHE, Sorbonne Universit\'e, Universit\'e de Paris, CNRS/IN2P3, Paris, France}
	\affiliation{IPSA-DRII,  Ivry-sur-Seine, France}
	
	\author{C.~McGrew}
	\affiliation{State University of New York at Stony Brook, Department of Physics and Astronomy, Stony Brook, New York, U.S.A.}

    \author{L.~Munteanu }
	\affiliation{IRFU, CEA, Universit\'e Paris-Saclay, Gif-sur-Yvette, France}

	\author{B.~Popov}
	\affiliation{LPNHE, Sorbonne Universit\'e, Universit\'e de Paris, CNRS/IN2P3, Paris, France}

   \author{D.~Sgalaberna }
	\affiliation{ETH Zurich, Institute for Particle Physics and Astrophysics, Zurich, Switzerland}
	
   \author{S.~Suvorov}
	\affiliation{LPNHE, Sorbonne Universit\'e, Universit\'e de Paris, CNRS/IN2P3, Paris, France}
	\affiliation{Institute for Nuclear Research of the Russian Academy of Sciences, Moscow, Russia}

  \author{X.~Y.~Zhao}
	\affiliation{ETH Zurich, Institute for Particle Physics and Astrophysics, Zurich, Switzerland}

\begin{abstract}
\noindent 
The most challenging and impactful uncertainties that future accelerator-based measurements of neutrino oscillations must overcome stem from our limited ability to model few-GeV neutrino-nucleus interactions. In particular, it is crucial to better understand the nuclear effects which can alter the final state topology and kinematics of neutrino interactions, inducing possible biases in neutrino energy reconstruction. The upgraded ND280 near detector of the T2K experiment will directly confront neutrino interaction uncertainties using a new suite of detectors with full polar angle acceptance, improved spatial resolutions, neutron detection capabilities and reduced tracking thresholds. In this manuscript we explore the physics sensitivity that can be expected from the upgraded detector, specifically focusing on the additional sensitivity to nuclear effects and how they can be constrained with future measurements of kinematic variables constructed using both outgoing lepton and nucleon kinematics. 

\end{abstract}

\maketitle

\section{Introduction}
\label{sec:introduction}

Neutrino oscillations are measured at accelerator-based experiments by inferring the rate of interactions of a particular neutrino flavour, as a function of neutrino energy, at a detector placed some distance from the neutrino production point (the ``far detector''). However, in order to meaningfully interpret far detector data in terms of neutrino oscillation probabilities, it is essential to well understand the unoscillated incoming neutrino flux, the detector response and the relevant neutrino interaction cross sections. It is also crucial to be able to estimate the bias and smearing of neutrino energy estimators due to ``nuclear-effects'' in neutrino interactions~\cite{Alvarez-Ruso:2017oui}. To constrain systematic uncertainties in oscillation measurements, neutrino-beam experiments therefore usually employ an additional detector placed close to the beam production point (the ``near detector''), before any oscillation is expected to have occurred. Measurements at the near detector are then used to constrain predictions at the far detector. 

\begin{figure}[hb]
\centering
\includegraphics[width=0.98\linewidth]{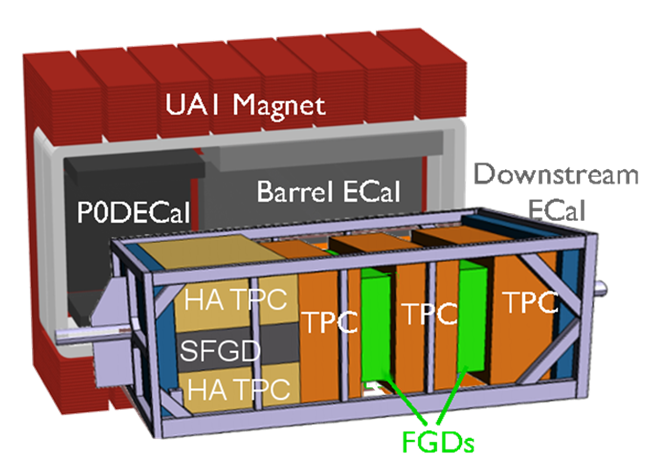}
\caption{\label{fig:ndup} An exploded view of the ND280 Upgrade detector. Neutrinos enter from the left of the image. Figure adapted from Refs~\cite{Abe:2011ks,T2K:2019bbb}.}
\end{figure}

The upgraded ND280 detector (``ND280 Upgrade'') considered in this paper is an improvement to the existing near detector of the T2K experiment~\cite{T2K:2011qtm} (``ND280''), which will also serve as a key component of the upcoming Hyper-Kamiokande (``Hyper-K'') experiment's~\cite{Hyper-Kamiokande:2018ofw} near detectors. T2K and Hyper-K are long-baseline experiments which measure neutrino oscillations over a distance of 295~km using intense muon neutrino and anti-neutrino beams with a neutrino energy peak at $\sim$600~MeV. The complete ND280 Upgrade detector consists of several sub-detectors encased in a magnet providing a uniform 0.2~T field, as shown in Fig.~\ref{fig:ndup}. The downstream sub-detectors of the current ND280 comprise a central tracking region consisting of ``fine-grained'' scintillator detectors (FGDs)~\cite{T2KND280FGD:2012umz} to act as targets for neutrino interactions, and low-density gaseous time projection chambers (TPCs)~\cite{T2KND280TPC:2010nnd} to track charged particles emitted in neutrino interactions, surrounded by an electromagnetic calorimeter (ECal). The new components of ND280 added in the upgrade are the two horizontal ``high angle'' TPCs (\hatpc) and the horizontal ``Super-FGD''~\cite{T2K:2019bbb}. The former are improved versions of the existing TPCs which will significant extend the detector's angular acceptance and resolution~\cite{Attie:2019hua}, whilst the latter is a new type of active target for neutrino interactions, made of more than 2 million 1~cm$^3$ optically isolated scintillating cubes connected with readout fibres in three orthogonal directions~\cite{Blondel:2017orl}. Six Time-of-Flight planes will be installed around the HA-TPCs and the \sfgd to verify whether particles crossing the \sfgd are leaving or entering the active volume~\cite{Korzenev:2019kud} and to provide complementary information for particle identification.

ND280 Upgrade is especially well adapted to characterise the aspects of neutrino interactions that are responsible for some of the most impactful and challenging sources of systematic uncertainties in T2K measurements. In particular, the 3D readout and fine segmentation of the \sfgd allows for an excellent hadron reconstruction with very low thresholds in a hydrocarbon material. This, in turn, allows a characterisation of the nuclear effects that can significantly bias neutrino energy reconstruction~\cite{Alvarez-Ruso:2017oui} and which were responsible for some of the largest uncertainties in T2K's latest neutrino oscillation analysis~\cite{T2K:2019bcf, T2K:2021xwb}. In addition, the Super-FGD's neutron detection capabilities~\cite{T2K:2019bbb} both extend its sensitivity to anti-neutrino interactions and allow a potential isolation of an almost nuclear-effect free sample of neutrino interactions on hydrogen~\cite{Munteanu:2019llq}.

The goal of this paper is to provide a quantitative estimate of ND280 Upgrade's sensitivity to the most important sources of systematic uncertainty in T2K and Hyper-K oscillation analyses, specifically through its ability to accurately reconstruct nucleons alongside the charged lepton in charged-current interactions with no mesons in the final state (the primary interaction topology for T2K and Hyper-K). These sensitivities are presented as a function of accumulated statistics, covering both the T2K and Hyper-K eras. The analysis method is presented in Sec.~\ref{sec:methodology}, where the simulation and the reconstruction is described, followed by a discussion of the strategy to extract meaningful sensitivities. The resultant sensitivities are then discussed in Sec.~\ref{sec:results} before conclusions are drawn in Sec.~\ref{sec:conclusion}.

\section{Methodology}
\label{sec:methodology}

\subsection{Simulation}
\label{sec:simulation}

Whilst the full simulation of the ND280 Upgrade is in active development within the T2K collaboration, the fundamental performances of the proposed detector technology has been extensively studied in various test-beams at CERN~\cite{Attie:2019hua, Blondel:2020hml}, LANL and DESY~\cite{Attie:2021yeh}. 
These data have allowed development of a preliminary \sfgd detector simulation and reconstruction framework. From this early simulation, the main detector effects (smearing and resolution) have been parameterised and applied to a sample of neutrino interactions generated with version 5.4.0 of the NEUT simulation~\cite{Hayato:2009zz} on a hydrocarbon scintillator (CH) target using the T2K neutrino and anti-neutrino flux predictions~\cite{Abe:2012av,t2kfluxurl}.

Within the reconstruction framework, charge deposits in the \sfgd are identified and processed to form tracks. For contained tracks, the charge deposits along most of the track's length are considered to identify the particle energy loss over the whole track range, permitting both momentum reconstruction and particle identification (PID). A small region very close to the interaction vertex is willingly neglected in order to negate the impact of energy deposition from other interaction products in the vicinity of the target nucleus. Tracks with an ambiguous particle identification and those which undergo apparent secondary interactions (i.e. from an observed deflection of peak in the charge deposits) are rejected\footnote{Further reconstruction efforts will aim to recover some of this lost selection efficiency}. It should be noted that the non-uniformity of energy deposits due to the position of particles within a scintillator cube is not yet modelled (which could alter the momentum resolution for short tracks by $\sim$10\%), but it has been ensured that such changes in resolution would not significantly affect the results presented here. Tracks which are not fully contained and enter a TPC are split into two parts: the \sfgd segment is reconstructed as described above and the TPC segment undergoes a parameterised reconstruction based on known performances of the current T2K TPC~\cite{T2KND280TPC:2010nnd}. The improved performance expected from the HA-TPCs is not yet taken into account (meaning the detector performance for the TPC segments of high angle tracks is expected to be slightly underestimated).

The input NEUT simulation uses the Spectral Function model~\cite{Benhar:1994hw} for Charged-Current Quasi-Elastic (CCQE) interactions, the Valencia model for a multi-nucleon meson exchange current contribution~\cite{Nieves:2011pp} and the Rein-Seghal model~\cite{Rein:1981ys} for single-pion production. The Deep-Inelastic-Scattering channel is simulated using the GRV98~\cite{GRV98} parton distribution functions with Bodek-Yang corrections~\cite{Bodek:2005de} for the cross section, whilst hadron production is modelled by both PYTHIA 5.72~\cite{SJOSTRAND199474} and a custom model~\cite{Aliaga:2020rqb}. It should be noted that the CCQE model contains both a mean-field and a ``short range correlation'' (SRC) component, the latter of which produces two outgoing nucleons. Final state interactions (FSI) of hadrons are described using cascade models tuned to hadron-nucleus scattering data~\cite{Hayato:2009zz, PinzonGuerra:2018rju}. This is the same neutrino interaction model that is used in the last T2K oscillation analysis and described in details in Ref.~\cite{T2K:2021xwb}. NUISANCE~\cite{Stowell:2016jfr} is used to process the NEUT output. In total 6 million neutrino interactions and 2 million anti-neutrino interactions are simulated, corresponding to an estimated 3.0$\times 10^{22}$ and 4.5$\times 10^{22}$ protons impinging on the target (POT) in the neutrino beamline respectively. For reference, T2K's latest analysis used 1.49~(1.64)$\times 10^{21}$ POT of neutrino (anti-neutrino) data whilst the Hyper-K design report~\cite{Hyper-Kamiokande:2018ofw} considered a total of 27$\times 10^{21}$ POT (in a 1:3 neutrino over anti-neutrino ratio), corresponding to 10 years data taking. The simulation is scaled to test sensitivity as a function of accumulated statistics assuming 1.9 tons of mass in the \sfgd fiducial volume. 

\begin{figure}[hb]
\centering
\includegraphics[width=0.98\linewidth]{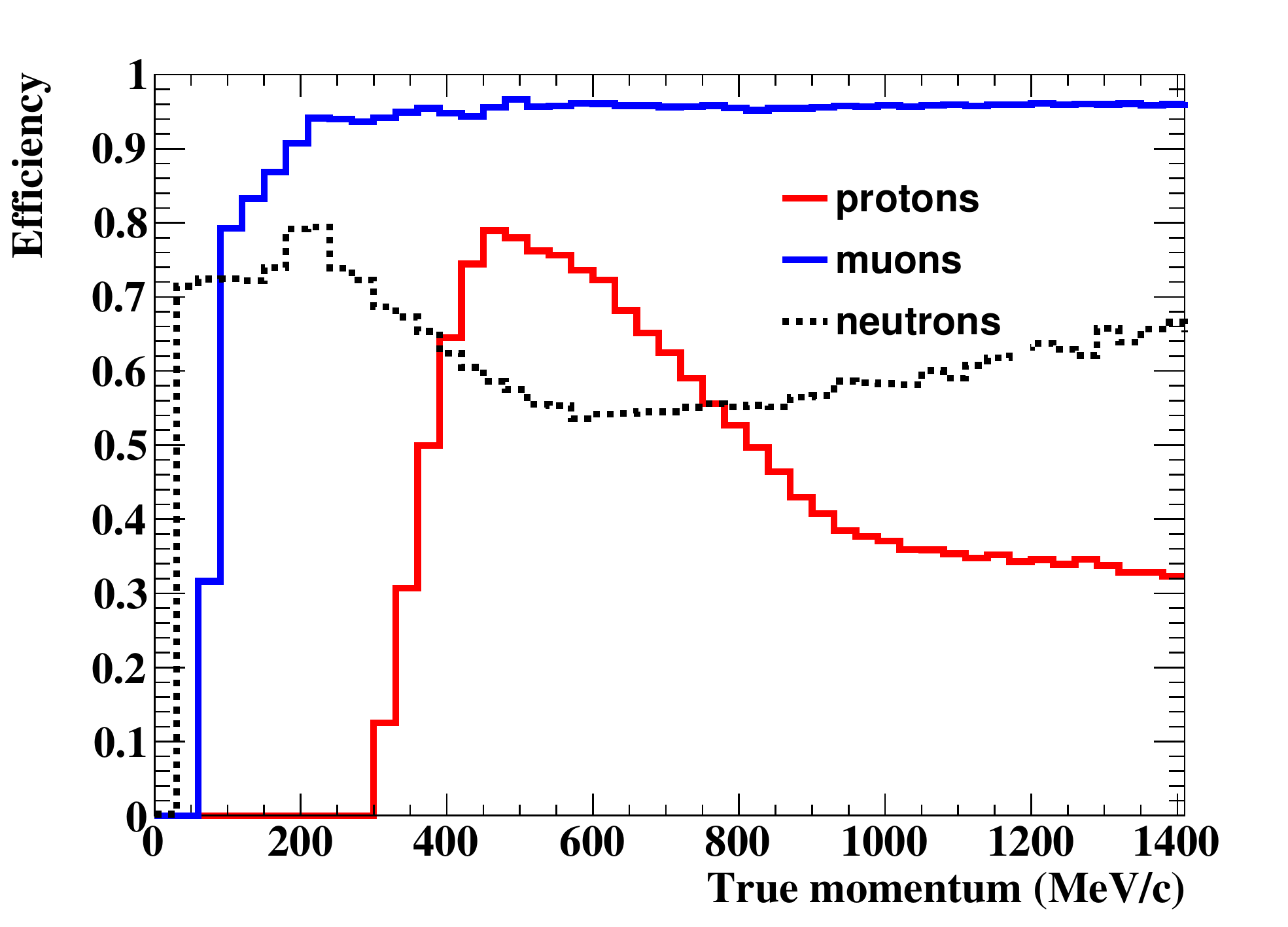}
\includegraphics[width=0.98\linewidth]{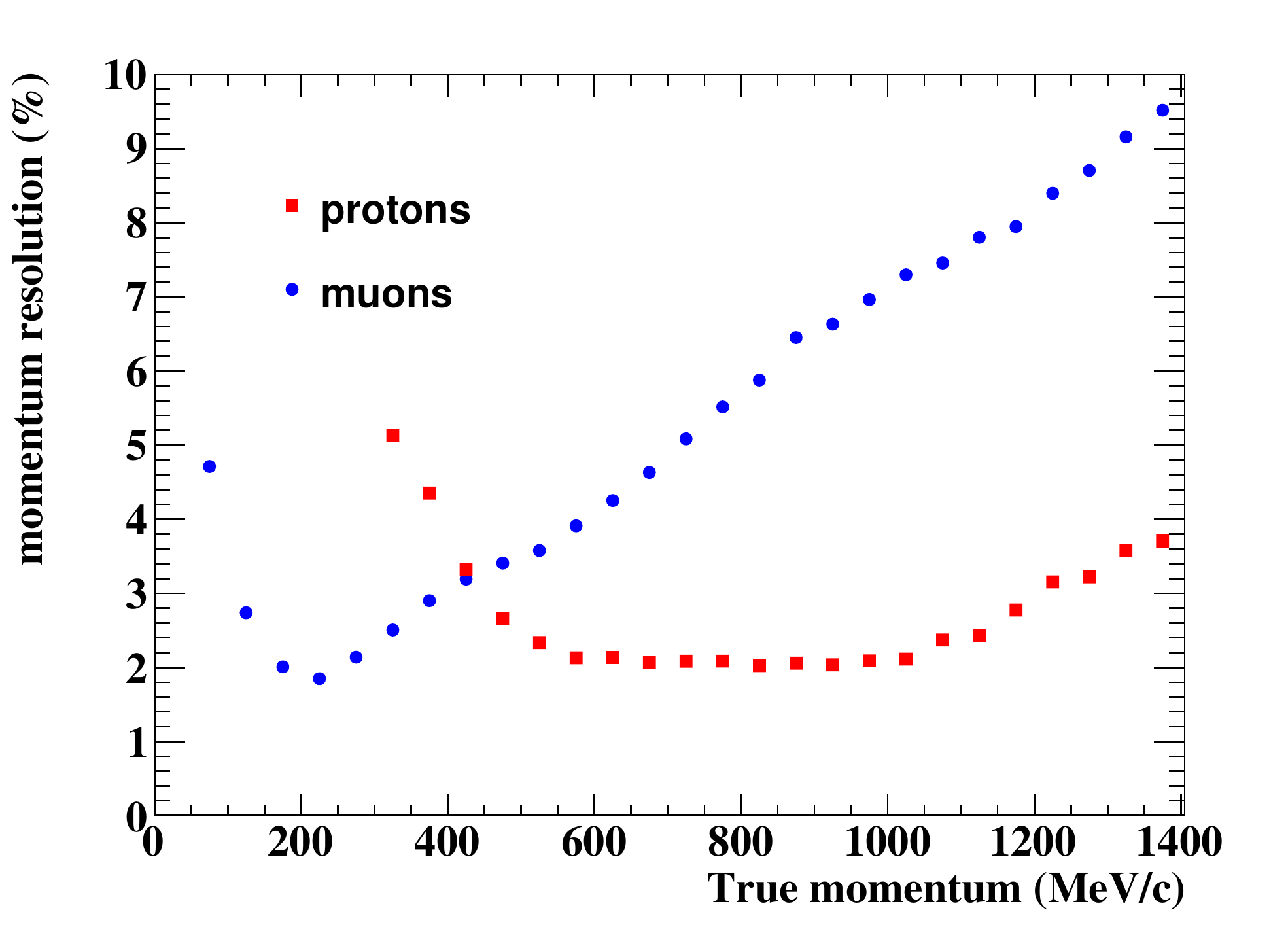}
\caption{\label{fig:simuCharged} \textbf{Top:} the efficiency to detect protons, muons and neutrons and \textbf{bottom:} the momentum resolution of muons and protons all as as a function of their respective momenta. The neutron momentum resolution ranges from $\sim$15\% to 30\% and is discussed in more detail in Ref.~\cite{Munteanu:2019llq}. }
\end{figure}

The parameterisation of detector effects is applied for protons, neutrons, muons and charged pions on a particle-by-particle basis. For charged particles a Gaussian momentum and angular smearing is applied alongside a probability to not reconstruct the particle (to model inefficiencies). These response functions are applied based on a particle's type and as a function of true momentum and direction. Neutron resolutions and efficiencies are also applied and are handled as described in Ref.~\cite{Munteanu:2019llq} (using the method where the time resolution depends on the light yield within a cube). In this analysis no cut is made on the distance the neutron travels from the interaction vertex (i.e. no ``lever-arm'' cut on the neutron propagation distance is applied), which increases neutrino detection efficiency but also degrades the momentum resolution. The modelled detector performance is summarised in Fig.~\ref{fig:simuCharged}, which describes the momentum resolution and selection efficiency for muons, protons and neutrons. The decrease in proton selection efficiency after 500~MeV/c is largely from track rejection due to identified secondary interactions, but at higher momentum ($>$~1~GeV/c) track rejection from ambiguous PID also plays a role. The parabolic shape of the resolutions stems from difficulties in reconstructing very short tracks, followed by peak performance for fully contained tracks, while higher momentum tracks reach the TPCs with relative resolution worsening at higher momentum (as expected due to the smaller curvature in the magnetic field).

Table~\ref{tab:events} shows the number of reconstructed events in a sample of Charged-Current interactions without reconstructed pions in the final state (CC0$\pi$) and with at least one reconstructed proton/neutron for neutrino/anti-neutrino interactions. In this study we consider only these CC0$\pi$ selections, which is the dominant interaction topology for T2K oscillation analyses. For reference, the current ND280 would expect to select $\sim$38,000 neutrino interactions with at least one reconstructed proton in the final state for $5\times 10^{21}$ POT~\cite{Abe:2018pwo} and has not been shown to be able to reconstruct neutrons.   

It should be noted that this simple parameterised approach can not account for all the physics of a full reconstruction and event selection. Most importantly neutral pions are assumed to be always rejected and the possibility of misidentifying one particle type as another is not considered beyond the aforementioned impact on the selection efficiency. Whilst these are important limitations, in general they are subdominant effects for the CC0$\pi$ selections considered in this paper. Current ND280 CC0$\pi$ selections with a reconstructed proton in the final state have around 20\% background fully dominated by undetected pions, only a small minority of this background stems from a mis-identification of particle type or from an undetected neutral pion~\cite{T2K:2021xwb, Abe:2018pwo}. 

\begin{table}[htbp]
\centering
\begin{tabular}{|l|c|c|}
\hline
\textbf{Channel} & \textbf{Neutrino events} & \textbf{Anti-neutrino events}  \\
\hline
CCQE & 149 911 & 54 875 \\
2p2h & 31 505 &  9 558\\
Pion absorption & 24 483 & 3 357 \\
Undetected pions & 18 100 & 2 914\\
\hline
 \end{tabular}
 
   \caption{\label{tab:events} Number of reconstructed events in CC0$\pi$ channels for $5\times 10^{21}$ Protons on Target in neutrino mode and $5\times10^{21}$ in anti-neutrino mode. 
   } 

\end{table}

\subsection{Analysis strategy}
\label{sec:analysis-strategy}

The T2K oscillation analysis (see e.g. Ref.~\cite{T2K:2021xwb}) relies on a fit of ND280 data to constrain a parameterised model of the neutrino flux and interaction cross sections. The primary ND280 samples are split depending on whether zero, one or more charged pions are reconstructed in the final state. The distribution of muon kinematics (momentum and angle) in each sample is then compared to the model, thereby providing constraints on the parameters. 
The results of applying the same analysis to a sample of events simulated in ND280 Upgrade are shown in Ref.~\cite{T2K:2019bbb}. In this paper a similar strategy is applied for the CC0$\pi$ channel only, but now fitting data presented as a function of kinematic variables including lepton and hadron kinematics (protons and neutrons for the CC0$\pi$ channel) which better exploits the performance assets of the \sfgd.

Various observables for characterising the hadronic part of the final state have been investigated in neutrino cross-section measurements at ND280~\cite{Abe:2018pwo}, and other experiments (e.g. MINERvA~\cite{Lu:2018stk}). These include measurements of proton multiplicity and kinematics as well as of correlations between outgoing protons and muons~\cite{Lu:2015tcr, Furmanski:2016wqo}, which are particularly sensitive to the nuclear effects that often drive the dominant uncertainties in neutrino oscillation analyses~\cite{Alvarez-Ruso:2017oui}. The improved performances of ND280 Upgrade in the measurements of such variables are shown in Ref.~\cite{T2K:2019bbb}. In this analysis a a fit to these variables is performed to estimate ND280 Upgrade's potential to constrain flux and neutrino-interaction systematic uncertainties for future oscillation analyses. 

It should be highlighted that the optimal analysis with ND280 Upgrade would be a multidimensional fit including the muon kinematics, the hadronic kinematics and their correlations. The tools to implement such an approach are under development within T2K. In this paper the study is limited to fitting pairs of variables at once, focusing only on those which exploit the lepton-nucleon correlations. The results are therefore highlighting only the {\it additional} sensitivity that such new variables could provide, with respect to the existing T2K analyses. 

\subsubsection{Fit variables}
\label{sec:variables}

\begin{figure}
\begin{center}
\includegraphics[width=0.48\textwidth]{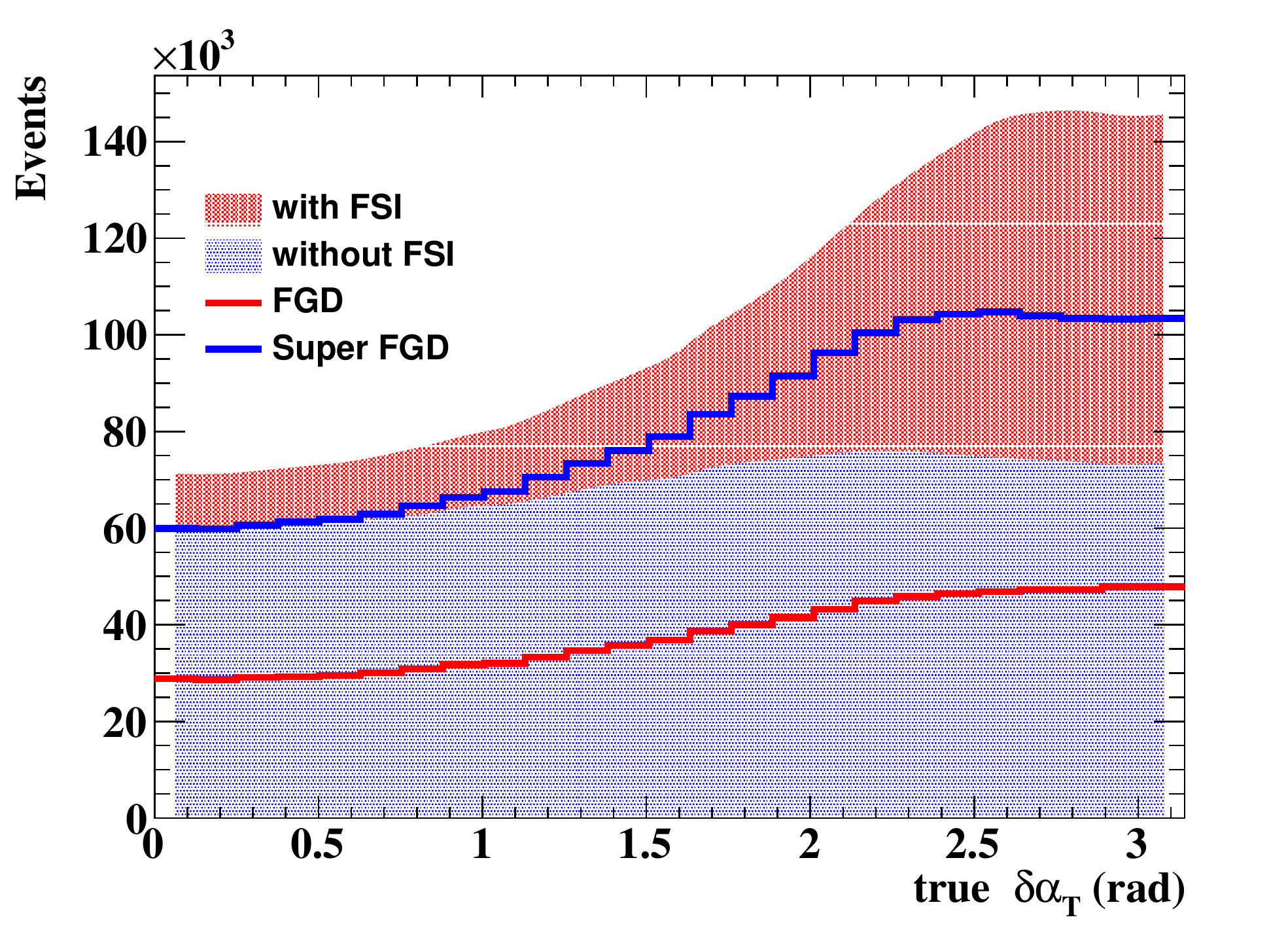}
\end{center}
\caption{The shaded regions show the generator level $\delta \alpha_T$ distribution for all CC0$\pi$ interactions for $3\times 10^{22}$ POT split by whether or not the outgoing nucleon underwent FSI. The overlaid solid lines indicate how the total $\delta \alpha_T$ distribution (including the with and without FSI components) changes when imposing the current ND280 FGD proton tracking threshold (450 MeV/c) and expectation from the \sfgd (300 MeV/c). 
}
\label{delta_alphaT}
\end{figure}

The inclusion of nucleon kinematics into the analysis requires a strategy to keep the nucleon FSI  under control. In future oscillation analyses this will act as a new nuisance uncertainty which 
must be understood in order to propagate constraints on other relevant nuclear effects from the near to far detectors.
The shape of the ``transverse boosting angle'', $\delta \alpha_T$, has been shown to be particularly sensitive to nucleon FSI~\cite{Lu:2015tcr} and is therefore chosen as one of the fit variables for this study. In Fig.~\ref{delta_alphaT}, the $\delta \alpha_T$ distribution for neutrino interactions is shown for CC0$\pi$ events with and without FSI. As shown, in the absence of FSI the $\delta \alpha_T$ distribution is expected to be almost flat whilst with FSI an enhancement at large $\delta \alpha_T$ is predicted (stemming from the fact that FSI tends to slow down the outgoing nucleons). Previous analyses of $\delta \alpha_T$ at ND280 have not been able to clearly identify this FSI induced enhancement because of its high proton tracking threshold ($\sim$ 450 MeV/c) which excludes many interactions in which the nucleon underwent FSI~\cite{Abe:2018pwo, Dolan:2018zye}. This is illustrated in Fig.~\ref{delta_alphaT}, which shows how the current ND280 tracking threshold flattens $\delta \alpha_T$ to a greater extent than that of ND280 Upgrade. It should also be noted that the neutron in anti-neutrino interactions can be reconstructed, through its time of flight, down to very low momentum threshold. This therefore potentially allows a useful FSI constraint even in the very low neutron momentum regions that cannot be measured for corresponding neutrino interactions~\cite{Munteanu:2019llq}.

\begin{figure}
\begin{center}
\includegraphics[width=0.48\textwidth]{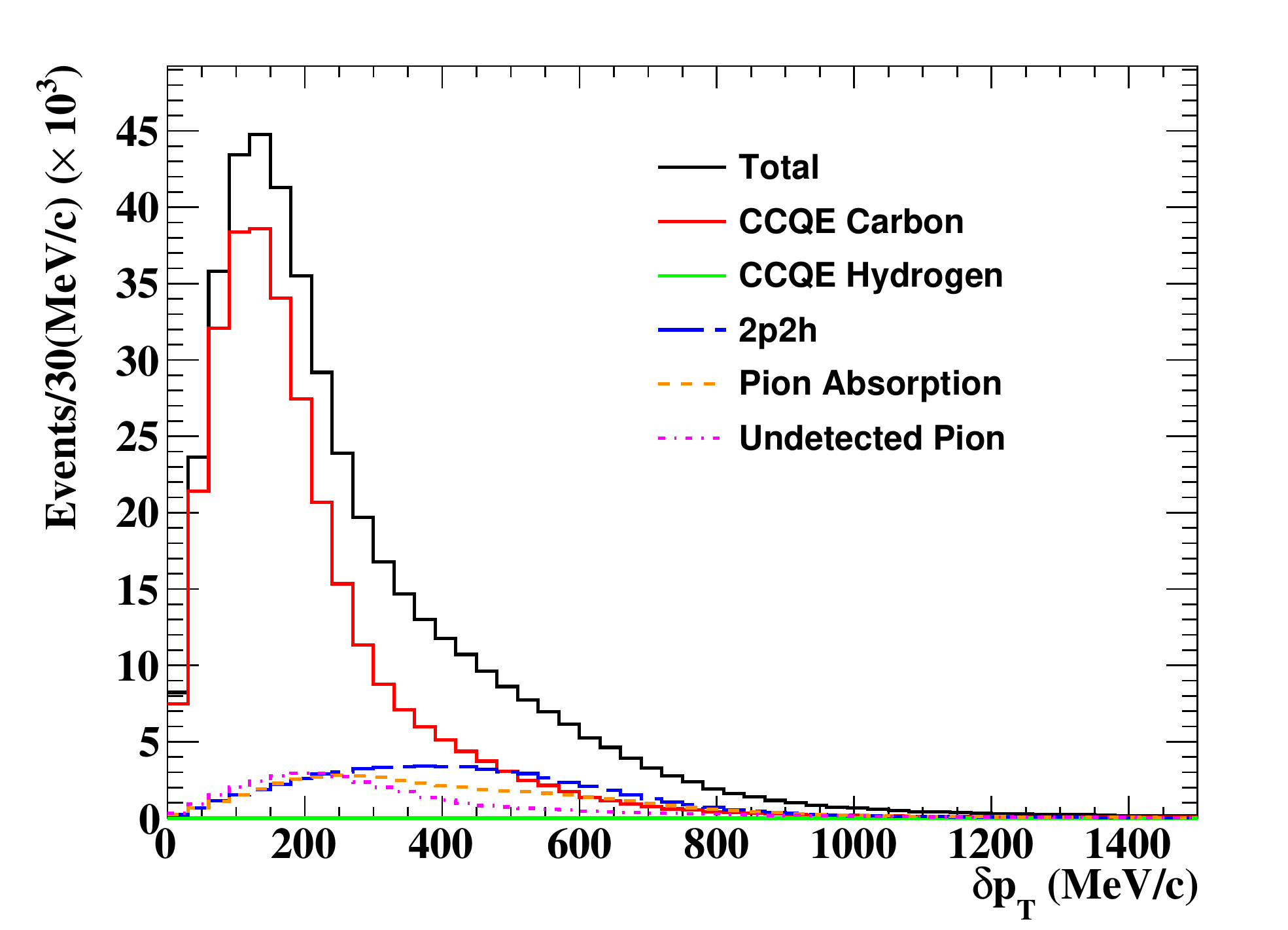}
\includegraphics[width=0.48\textwidth]{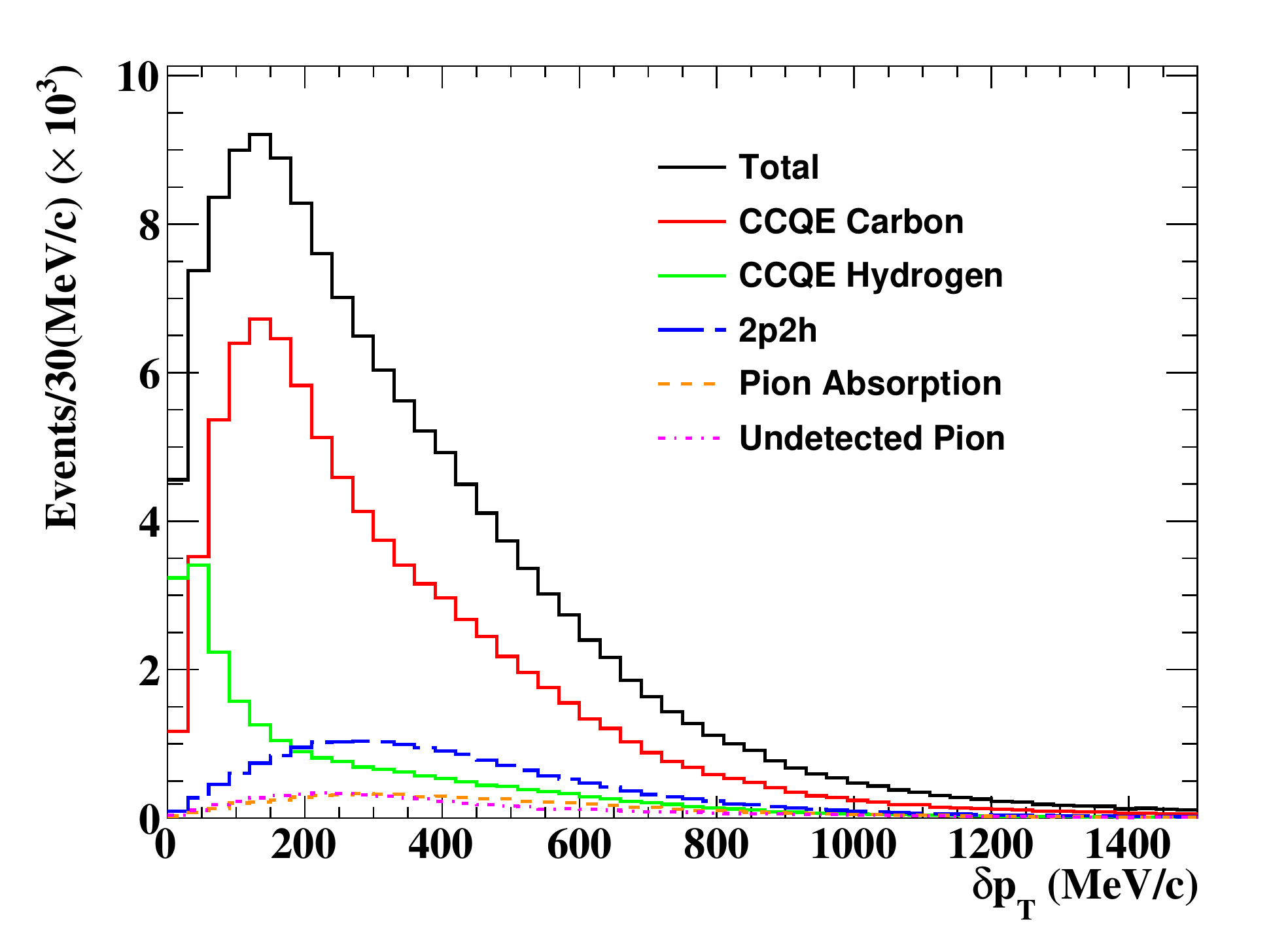}
\end{center}
\caption{The reconstructed $\delta p_T$ distribution for selected CC0$\pi$ neutrino (\textit{upper}) and anti-neutrino (\textit{lower}) interactions split by interaction mode and target for $1 \times 10^{22}$ POT.}
\label{dpt_all_mode_anu}
\end{figure}

Some of the most important systematic uncertainties for CCQE interactions in neutrino oscillation analyses are those which cause a bias in estimators for reconstructing neutrino energy. Such a bias directly impacts the measurement of neutrino oscillation parameters, notably $\Delta m^2_{32}$ and $\delta_{CP}$ precision measurements, especially in the presence of large CP violation. The principle cause of this bias stems from: the modelling of the initial target nucleon momentum within the nucleus (Fermi motion); the contribution of CC-non-QE or SRC processes entering CC0$\pi$ samples; and the ``removal'' energy it takes to liberate target nucleons from the nucleus (the latter was responsible for a dominant systematic uncertainty in T2K's latest neutrino oscillation analysis~\cite{T2K:2021xwb}). $\delta p_T$~\cite{Lu:2015tcr} and $p_N$~\cite{Furmanski:2016wqo}\footnote{Note that a subscript ``N'' (denoting nucleon) is used rather than ``n'' (for neutron) since this in these studies the variable can be calculated for neutrino or anti-neutrino interactions corresponding to probes of the initial state neutron or proton respectively.} are known to well separate CCQE from CC-non-QE in CC0$\pi$ samples whilst also well characterising the Fermi motion. In anti-neutrino interactions these variables can also separate the hydrogen and carbon contributions of the CH scintillator, thereby allowing some lifting of the degeneracy between nucleon and nuclear level effects, whilst also providing the opportunity for improved in-situ flux constraints~\cite{Munteanu:2019llq}. The reconstructed distribution of $\delta p_T$ for neutrino and anti-neutrino interactions are shown in Fig.~\ref{dpt_all_mode_anu}. The neutrino case demonstrates the clear separation of CCQE in the bulk and CC-non-QE in the tail (the small CCQE contribution to the tail is from SRCs and nucleon FSI). The anti-neutrino distribution does not show such good mode separation (due to the relatively poor neutron momentum resolution) but the shape difference of the hydrogen and carbon contributions is clearly visible.

Whilst variables characterising transverse kinematic imbalance are sensitive to many of the most important systematic uncertainties for neutrino oscillation analyses, they are not particularly sensitive to nuclear removal energy effects. A constraint can instead be established by looking for systematic shifts from expectation in the reconstructed neutrino energy distribution. However, in order to strongly constrain the removal energy uncertainty, a very good resolution is necessary in the reconstruction of neutrino energy at the near detector, together with a very good control of energy scale in the detector and of the flux energy peak. 

The usual estimator of neutrino energy is based on the very well known (e.g. see Ref.~\cite{T2K:2021xwb}) formula for CCQE events, relying on muon kinematics only:
\begin{align}
    E_{QE} = \dfrac{m_p^2-m_\mu^2-(m_n-E_{B})^2+2E_\mu(m_n-E_{B})}{2(m_n-E_{B}-E_\mu+p_\mu^z)},
\end{align}
where $m_{p/\mu/n}$ is the mass of a proton/muon/neutron; $E_\mu$ and $p_\mu^z$ is the outgoing muon energy and momentum in projected along the direction of the incoming neutrino; and $E_{B}$ is some assumed fixed nuclear binding energy of the struck nucleon (which is related to, but not exactly, the removal energy and is usually taken to be $\sim$25 MeV for carbon). 
A second estimator can be defined as:
\begin{align}
    E_{vis} = E_{\mu} +T_N,
\end{align}
where $T_N$ is the kinetic energy of the outgoing proton (neutron) in neutrino (anti-neutrino) interactions, $E_{\mu}$ is the total energy of the outgoing muon. $E_{vis}$ is the total visible energy of all outgoing particles in CC0$\pi$ events with one nucleon in the final state. Such an estimator, before detector smearing, is expected to be slightly smaller than the true neutrino energy in CCQE events due to the need to overcome the nuclear removal energy and the loss of energy through nucleon FSI. Similarly multinucleon interactions and pion absorption will populate the tail of low $E_{vis}$ since the second nucleon or the absorbed pion carry away some of the initial neutrino energy.

\begin{figure}
\begin{center}
\includegraphics[width=0.48\textwidth]{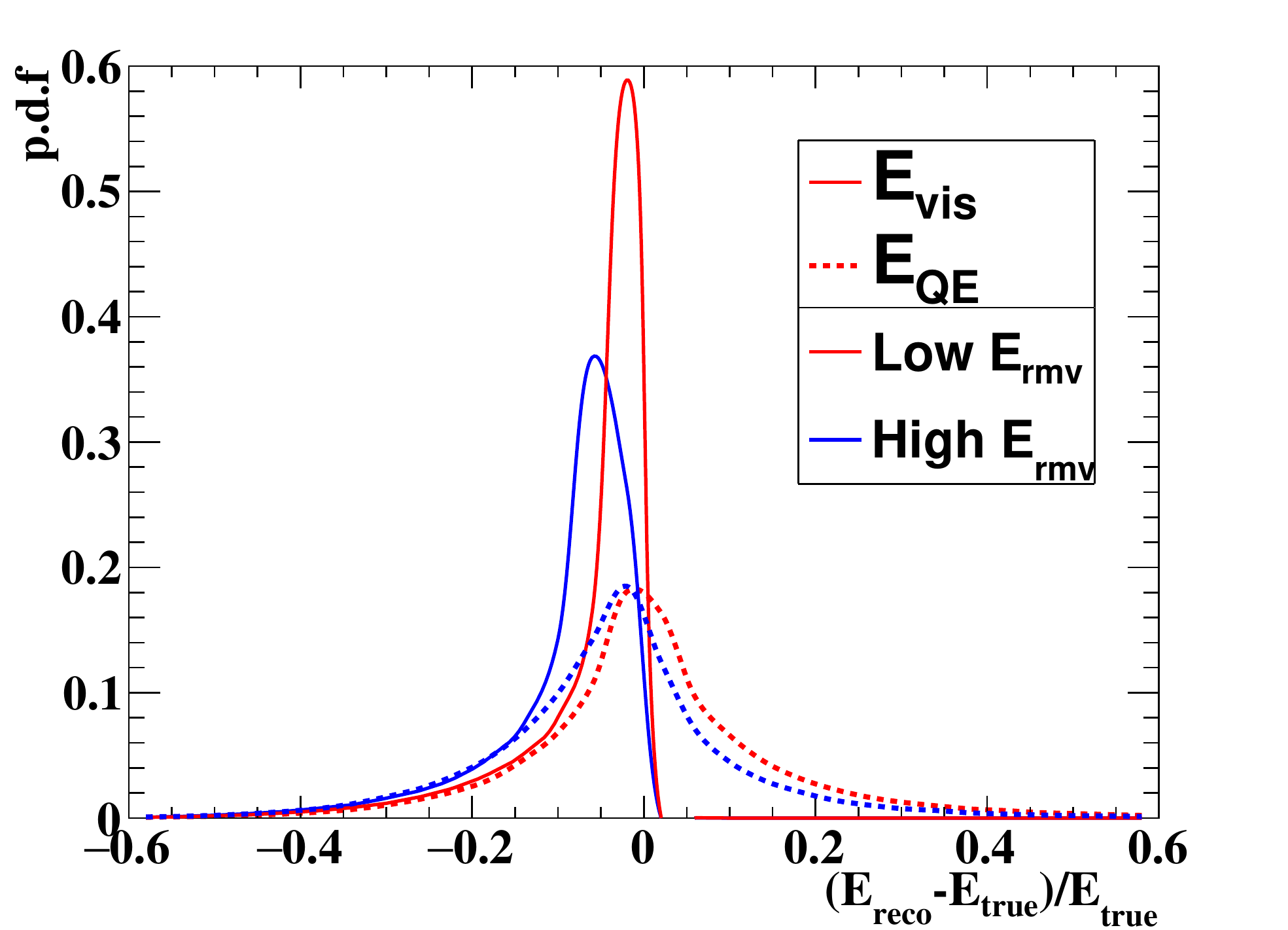}
\includegraphics[width=0.48\textwidth]{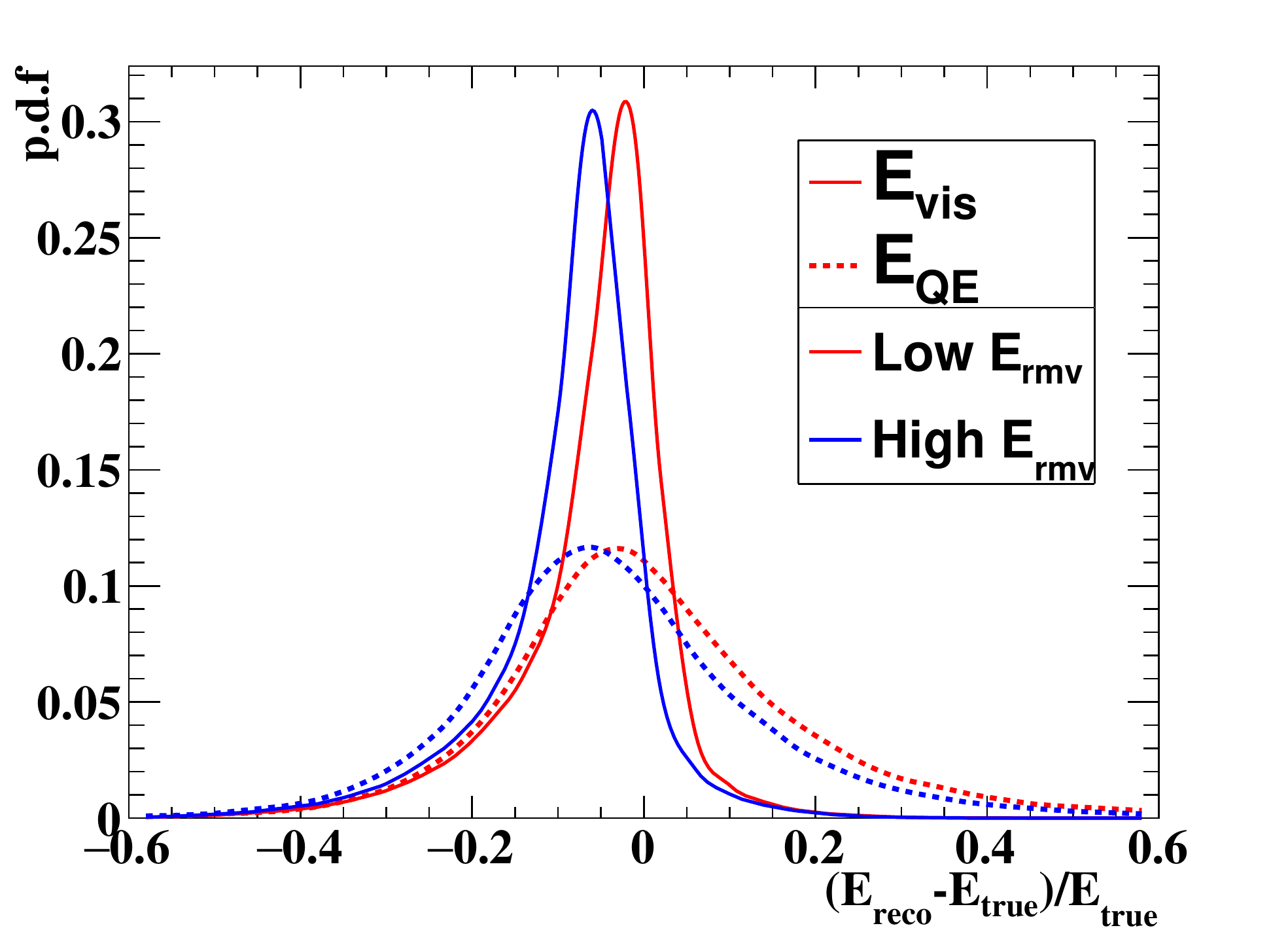}
\end{center}
\caption{The neutrino energy reconstruction resolution and bias is shown for the two estimators defined in the text ($E_{vis}$ and $E_{QE}$ as solid and dashed lines respectively) for $\pm 10$ MeV shifts to the nominal removal energy ($E_{rmv}$) distribution (denoted by the red and blue colours respectively). The upper plot does not include the effect of detector smearing such that $E_{reco}$ is constructed using the true muon and proton kinematics directly from the generator. In the lower plot $E_{reco}$ is instead built from the corresponding reconstructed quantities (i.e. with the detector smearing applied). 
}

\label{Evis_EnuQE}
\end{figure}

Fig.~\ref{Evis_EnuQE} compares the neutrino energy resolution for the two estimators alongside the impact of a possible bias due to removal energy. The distributions at generator level and after detector effects are shown. The $E_{vis}$ estimator has a higher peak at very good resolution ($< 5$\%) thus showing an increased sensitivity to possible bias in the removal energy estimation. This feature is preserved at reconstructed level, despite a larger experimental smearing of $E_{vis}$, due to the inclusion of proton tracking resolution in addition to the muon one. The observed smearing of $E_{vis}$ in data will be therefore mostly due to detector effects, while the smearing of $E_{QE}$ is dominated by Fermi motion. The smearing induced by tracking resolution can in principle be improved with more performant detectors. Moreover we expect to be able to model quite precisely the detector-induced smearing, thanks to test beam studies and detector simulation, while the smearing due to nuclear effects is typically less well known. The inclusion of both energy estimators and the study of their correlation may also enable future enhanced sensitivity to removal energy. 

To achieve a good sensitivity whilst keeping cross-section systematic uncertainties manageable, the pairs of variables used in the fit are one of the ``single transverse variables'', $\delta p_T$ or $\delta\alpha_T$, alongside the total visible energy (sum of the muon and nucleon energy in the final state). The full set of input histograms are shown in Appendix~\ref{app:allInputs}. The reconstructed Fermi momentum of the initial nucleon ($p_N$)~\cite{Furmanski:2016wqo} is also considered as an alternative to $\delta p_T$ and is discussed in Appendix~\ref{app:pn}.

\subsubsection{Fitter details}
\label{sec:details}

A binned likelihood fitter is built in two-dimensions, including statistical and systematic uncertainties. The statistical uncertainty is implemented as a Poisson term in the likelihood whilst the systematic uncertainties are parametrized as a function of the fit variables and implemented as nuisances with priors included mostly as Gaussian penalty terms in the likelihood. The prior uncertainties and fit variables are discussed in Sec.~\ref{sec:systematic}. One exception to the treatment of the prior uncertainties is an \textit{ad-hoc} ``uncorrelated uncertainty'' (also detailed in Sec.~\ref{sec:systematic}) which is added directly to the likelihood using the Barlow-Beeston approach~\cite{Barlow:1993dm}. The final $\chi^2$ used in the fitter is therefore defined as:

\begin{equation}
\begin{split}
    \chi^2=&\sum_{i=1}^{n\_bins}2\left(\beta_i E_i - O_i + O_i ln\frac{O_i}{\beta_i E_i} + \frac{(\beta_i-1)^2}{2\delta^2} \right) + \\
    &\sum_{j}\left(\dfrac{p'_{j}-p_{j}}{\sigma_j}\right)^2, \\
    \beta_j =& \frac{1}{2} \left(-(E_i \delta^2 - 1) + \sqrt{(E_i \delta^2 - 1)^2 + 4 O_i \delta^2} \right),
\end{split}
\end{equation}

where the first term is from the Poisson likelihood (with the Barlow-Beeston extension) and the second term is the Gaussian penalty. The definition of the Barlow-Beeston scaling parameter $\beta$ is also given. $O_i$ and $E_i$ are the observed and expected number of events for bin $i$, $\delta$ is the size of the uncorrelated uncertainty included directly in the Poisson likelihood. The second term is a sum over the systematic parameters in the fit where $p'_j$ and $p_j$ are the value of the parameter and its prior value respectively. $\sigma_j$ is the prior uncertainty of parameter $j$.

\subsubsection{Fit Parameters}
\label{sec:systematic}

The uncertainty model used in this analysis is designed to offer theory-driven conservative freedoms to modify pertinent aspects of the neutrino interaction model in addition to accounting for flux modelling and detector performance uncertainties. Particular care is taken to allow plausible variations of nucleon kinematics which are especially sensitive to nuclear effects. While this sensitivity is at the core of the improvements driven by ND280 Upgrade, it also requires a parametrisation of nuclear uncertainties beyond what is used by the latest T2K oscillation analyses, which exploit only measurements of lepton kinematics.

For NEUT's CCQE interactions the initial nuclear state is characterised by a two-dimensional spectral function (SF) which describes the momentum of initial state nucleons and the energy that it is necessary to remove them from the nucleus. This SF is broadly split into two parts: a part described by mean-field physics which has a shell structure (for carbon: one sharp p-shell and a diffuse s-shell) and a SRC part which results in two nucleons in the final state. The uncertainty model used allows a variation of the normalisation of each of the two mean-field shells separately and of the total strength of the SRC component. A wide prior uncertainty of 30\% is applied to the SRC and shell parameters. Further freedom is added through an overall removal energy shift parameter (similar to what is described in Ref.~\cite{T2K:2021xwb}) which is built using an interpolation between two alternative simulation histograms that are generated with opposite 10 MeV shifts of the removal energy. The removal parameter is not constrained with any prior uncertainty. For anti-neutrino interactions off a hydrogen target nuclear effects are not relevant, but a 5\% normalisation uncertainty is considered to account for imperfectly modelled nucleon-level effects. 



The 2p2h uncertainties are provided through two normalisation factors, one for $E_{vis} < 600$ MeV and one for the remaining phase space. This accounts for uncertainties in 2p2h nucleon ejection model, allowing 2p2h shape to change as a function of neutrino energy and a consistent treatment between fits with different pairs of variables. The 2p2h uncertainties are not constrained prior to the fit. 



Since the pion production contribution to the selected samples is low, uncertainties on this are modelled via two parameters: one controlling the normalisation of the pion production background (i.e. events in which a pion was not detected) and another to alter the normalisation of the pion absorption FSI contribution. For a conservative approach, neither are constrained with prior uncertainties. 

Nucleon FSI is treated similarly to the removal energy shift parameter, relying on an interpolation between histograms generated with NEUT using different nucleon FSI strengths. This is set via a 30\% modifications of nucleon mean free paths inside the FSI cascade to cover differences between different FSI models and the variation in nuclear transparency data~\cite{Niewczas:2019fro, Dytman:2021ohr}.

The flux uncertainties are handled using the T2K flux covariance matrix shown in Ref~\cite{Abe:2012av}, which provides correlated uncertainties on the flux between ranges of neutrino energy. Flux shape uncertainties are a second order effect, and so for computational ease the covariance is applied to bins of visible energy rather than true neutrino energy. In this way only flux uncertainties covering energies between 0.2 and 1.5 GeV are included. The range starts from 0.2 GeV since this is the starting point of $E_{vis}$ histogram, but this covers the ND280 flux peak and the primary region of interest for neutrino oscillation analysis. 


We additionally consider an ``uncorrelated uncertainty'' which accounts for systematic uncertainties in detector response modelling and small shape freedoms not included via the other parameters. Rather than existing as a fit parameter, this uncertainty is included by directly modifying the likelihood to effectively add an uncorrelated additional uncertainty to every bin of the analysis. Based on the size of detector uncertainties and background model shape parameters in T2K measurements of $\delta p_T$~\cite{Abe:2018pwo}, this is chosen to be 11.6\% at 6 $\times$ 10$^{21}$ POT and is then scaled up or down with the square root of the POT to represent improved constraints on the corresponding systematic uncertainties. 

The prior uncertainties in the systematic penalty term outlined in Sec~\ref{sec:analysis-strategy} and the uncorrelated uncertainty are summarised in Tab.~\ref{tab:priors}.

\begin{table}[htbp]
\footnotesize
\centering
\begin{tabular}{|l|c|c|}
\hline
\textbf{Parameter} & \textbf{Prior Constraint} & \textbf{Notes}  \\
\hline
p-shell norm. & 30\% &  \\
\hline
s-shell norm. & 30\% &  \\
\hline
SRC strength & 30\% &  \\
\hline
total QE normalisation & 10\% & \\
\hline
Removal energy shift & Unconstrained & \\
\hline
2p2h, low & Unconstrained & $<$600 MeV \\
\hline
2p2h, high & Unconstrained & $>$600 MeV\\
\hline
Undetected pions & Unconstrained &  \\
\hline
Pion FSI contribution & Unconstrained &  \\
\hline
Nucleon FSI strength & 30\%  &  \\
\hline
Flux (binned $E_{vis}$) & T2K cov.  & \\
\hline
Hydrogen normalisation & 5\%  & $\Bar{\nu}$ only \\
\hline
Uncorrelated Uncertainty & 11.6\% & No parameter fit,\\
 & (at 6 $\times$ 10$^{21}$ POT) & POT dependence \\
\hline
\end{tabular}
\caption{\label{tab:priors} A list of fit parameters, their prior constraints and notes regarding their application. Whilst not a fit parameter, the uncorrelated uncertainty is also listed. }
\end{table}

\section{Results and Discussion}
\label{sec:results}

\begin{figure}[h]
\begin{center}
\includegraphics[scale=0.45]{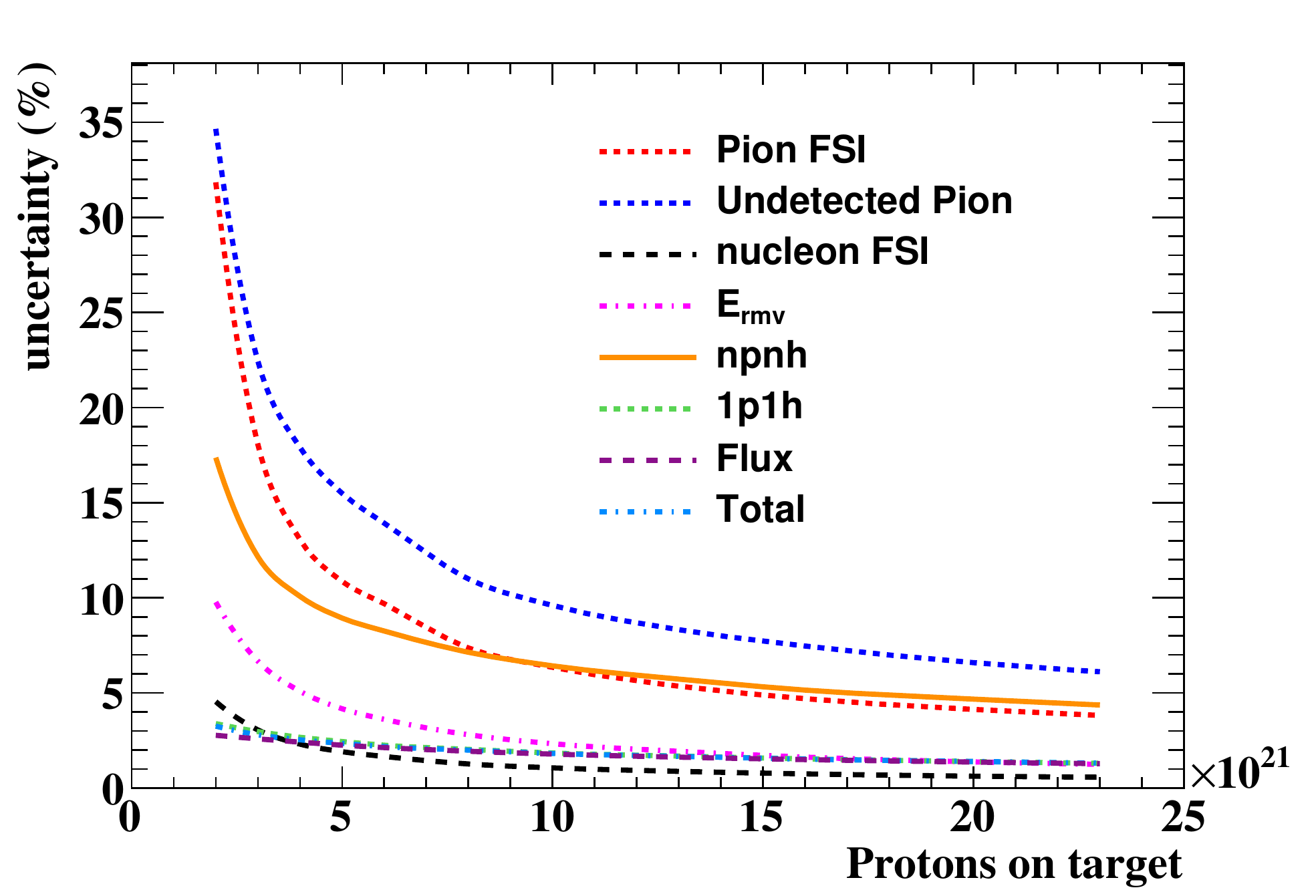}
\includegraphics[scale=0.45]{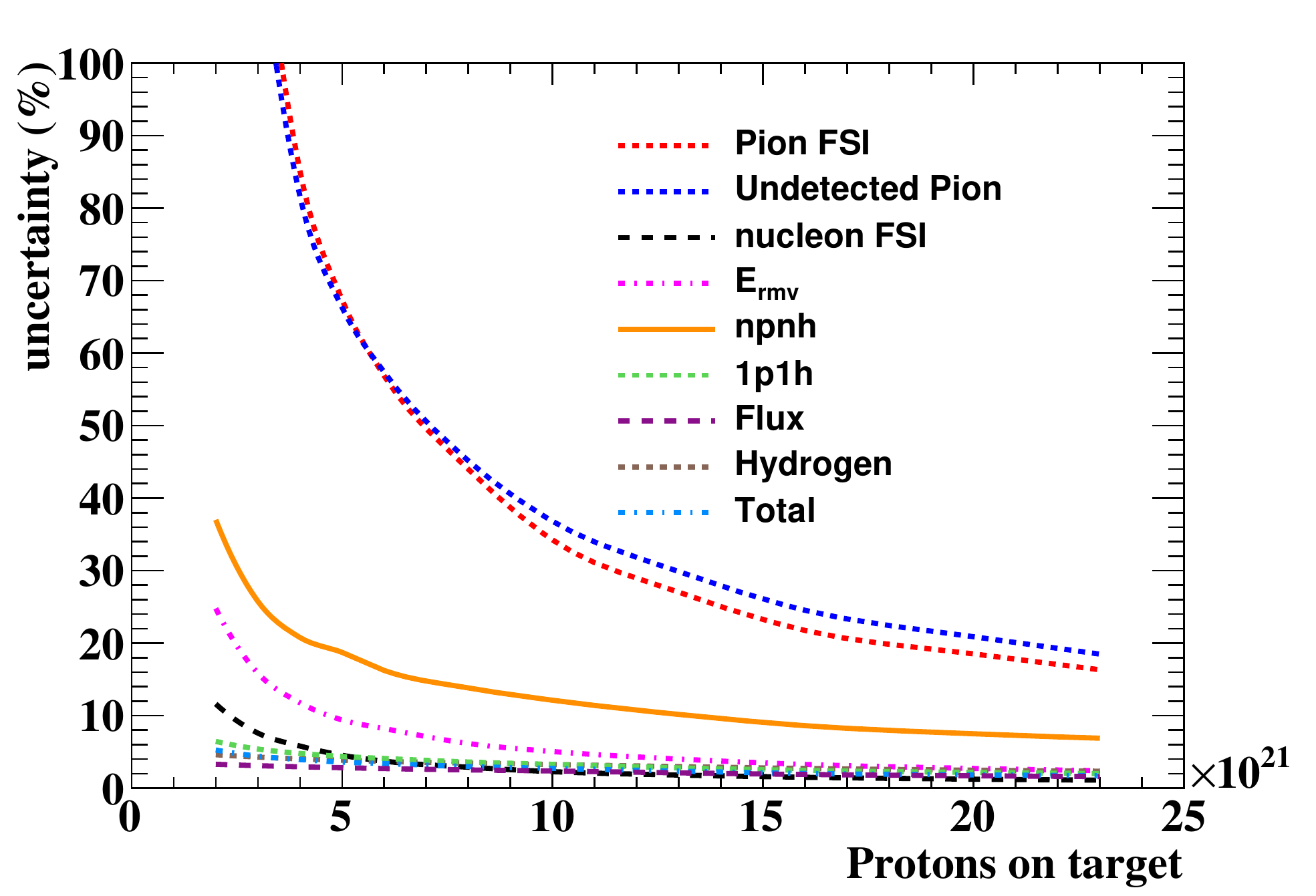}
\end{center}
\caption{The 1$\sigma$ sensitivity to systematic parameters as function of POT in neutrino case (top) and anti-neutrino case (bottom) when fitting the reconstructed CC0$\pi$ data binned in $\delta p_T$ and $E_{vis}$. The values in the plot are the ratio of the parameter uncertainty to the parameter nominal value. }
\label{fig:all_para}
\end{figure}

\begin{figure}
\begin{center}
\includegraphics[width=0.45\textwidth]{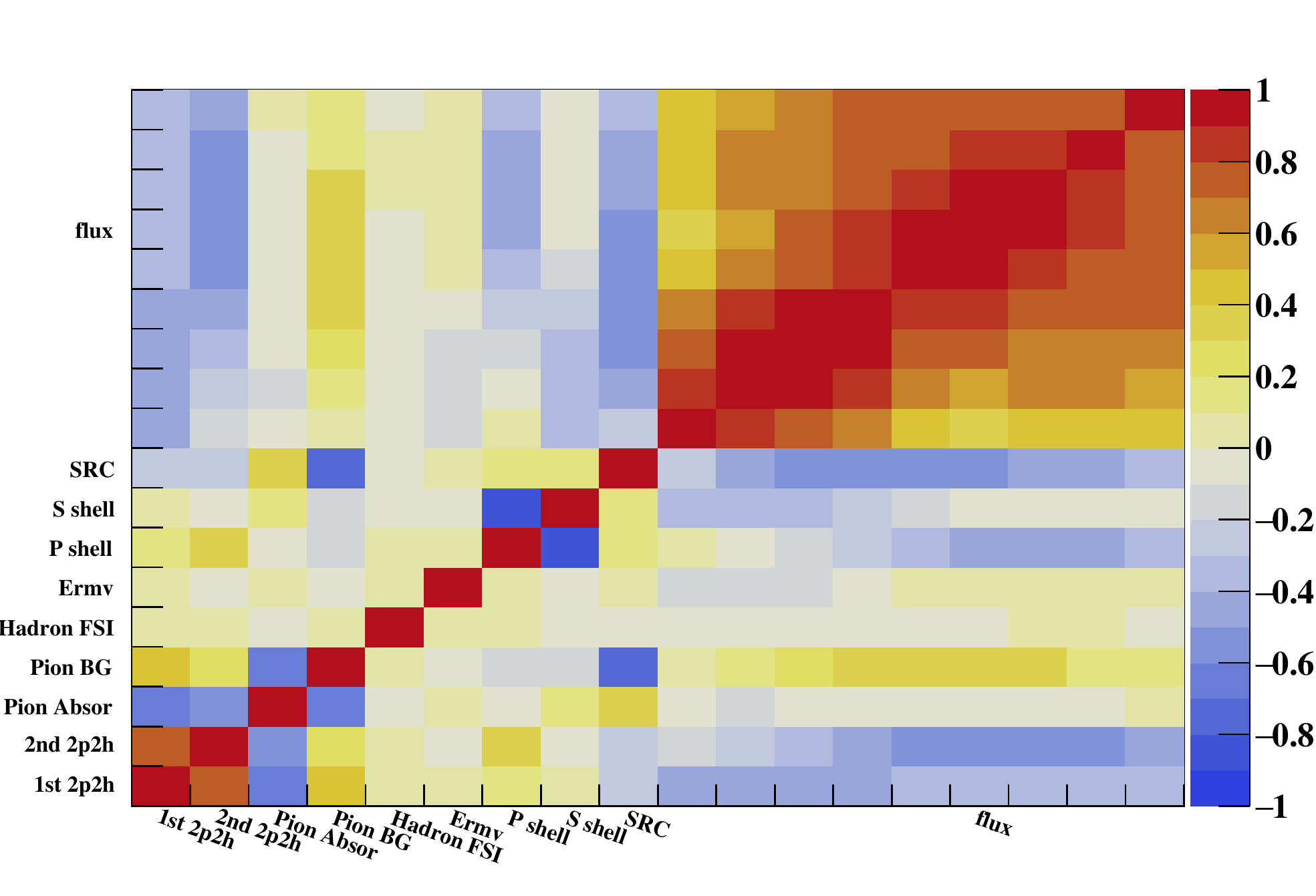}
\includegraphics[width=0.45\textwidth]{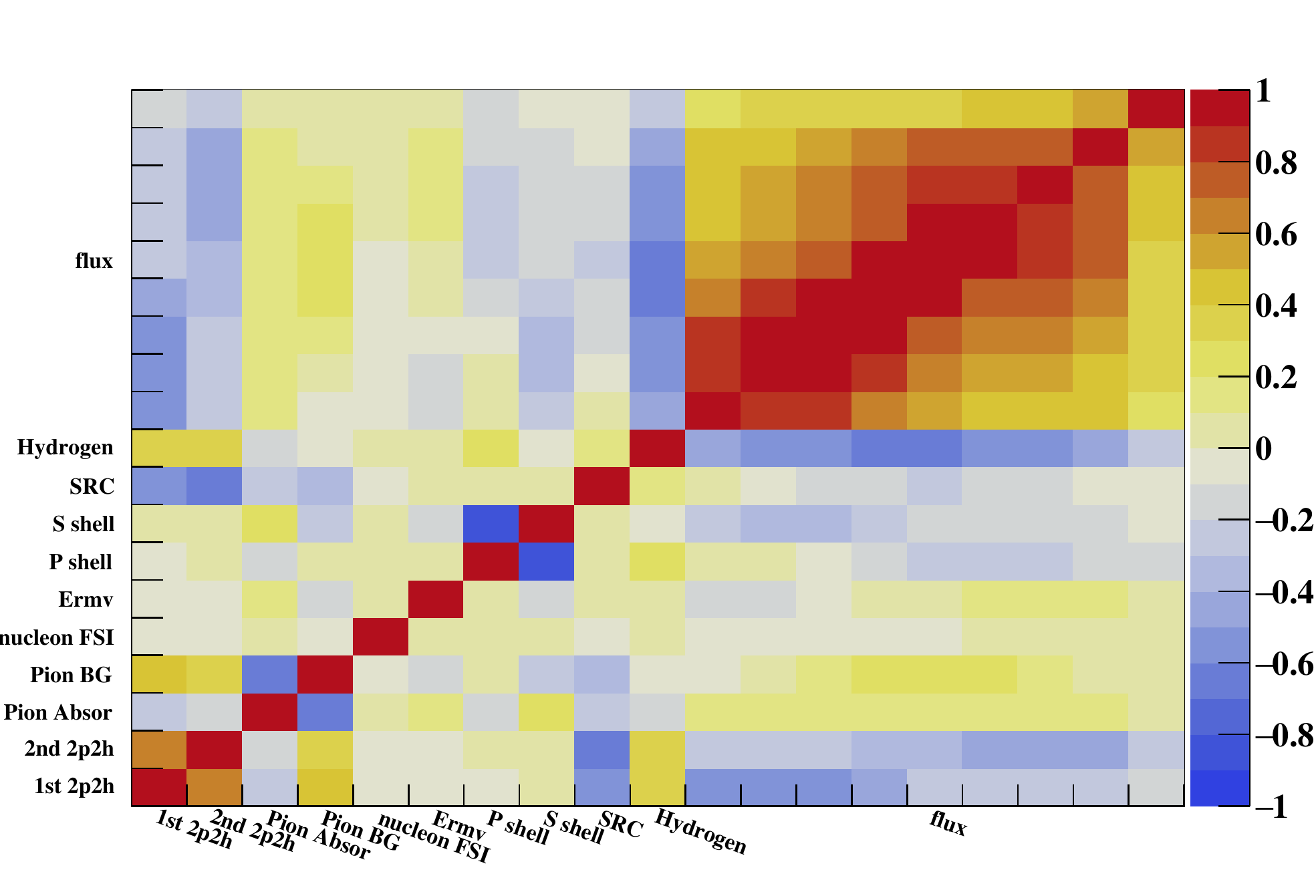}
\end{center}
\caption{The correlation matrix between constrained parameters following a fit to reconstructed CC0$\pi$ data binned in $\delta p_T$ and $E_{vis}$ with $1 \times 10^{22}$ POT. The results following a fit to neutrino (top) and anti-neutrino (bottom) samples are shown.}
\label{fig:CorrMat}
\end{figure}

The fit described in Sec.~\ref{sec:methodology} is performed and the parameter uncertainties obtained are evaluated as a function of the statistics accumulated (denoted by the simulated POT exposure). Note that all the systematic uncertainties are always fit together, even if the uncertainty on only one parameter is shown. A summary of the results are shown in Figs.~\ref{fig:all_para}~and~\ref{fig:CorrMat}, which shows parameter constraints as a function of POT and the correlations between fit parameters (for $10^{22}$ POT) respectively. In the former figure the parameters describing CCQE and 2p2h interactions are integrated to provide more meaningful uncertainties on the total one-particle-one-hole (1p1h) final state normalisation (CCQE without SRCs) and an accompanying npnh final state normalisation (2p2h and CCQE with SRCs). SRCs are combined with 2p2h since they have similar kinematic properties (as can be seen by their anti-correlation) and both are responsible for similar neutrino energy reconstruction bias. The uncertainty on the total cross section (integrating all cross-section systematic parameters but not the flux) is also shown. Tabulated sensitivities are shown for two fixed POT values in Tab.~\ref{tab:para_uncert1}.

\begin{figure}[h]
\begin{center}
\includegraphics[width=0.45\textwidth]{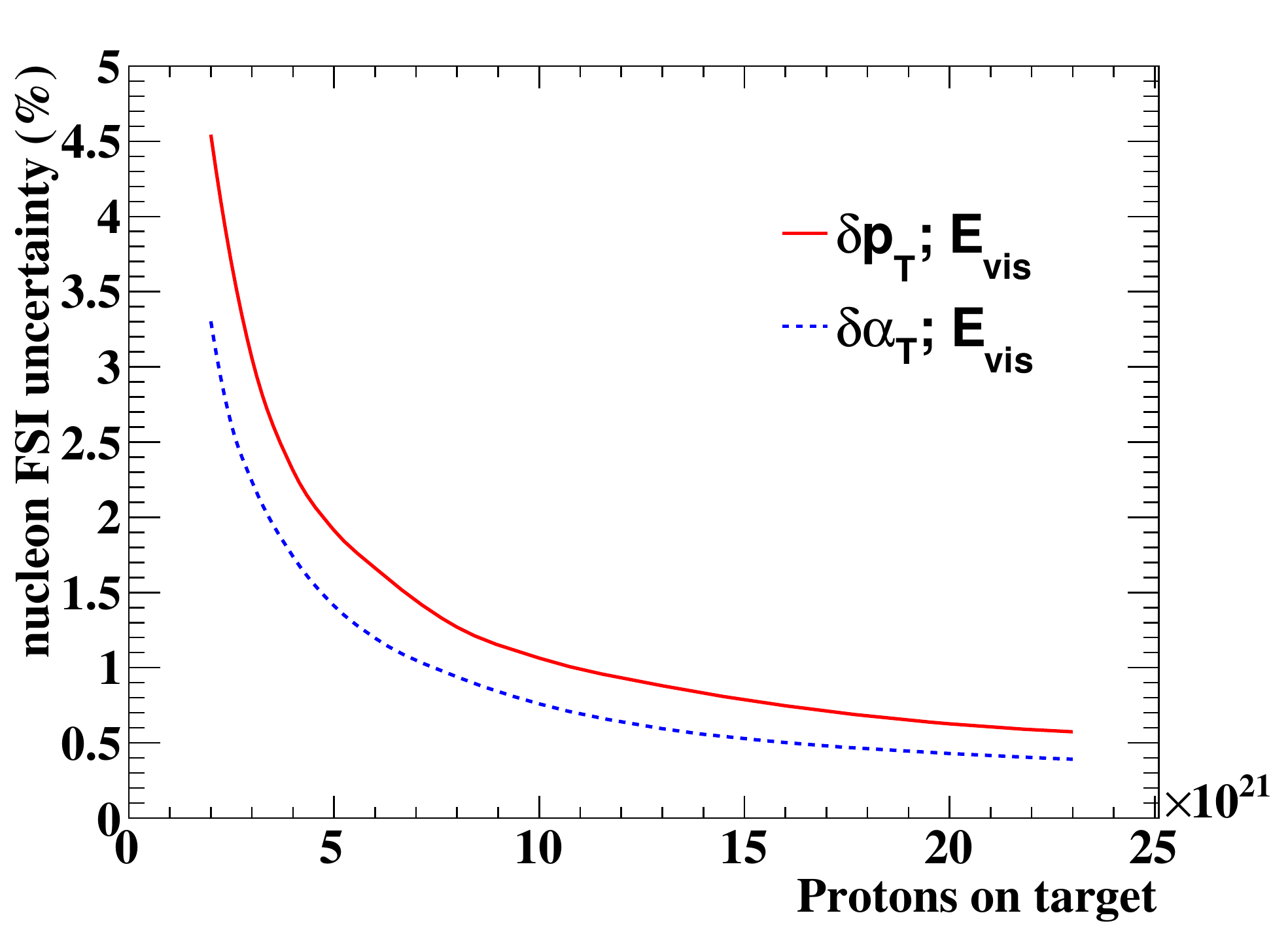}
\end{center}
\caption{The 1$\sigma$ sensitivity to the nucleon FSI parameter as a function of POT for neutrino interactions when fitting the reconstructed CC0$\pi$ data binned in $\delta p_T$ and $E_{vis}$ or in $\delta \alpha_T$ and $E_{vis}$.}
\label{protonFSI_postfit}
\end{figure}

As discussed in Sec.~\ref{sec:variables}, it is expected that the \sfgd should allow a particularly strong constraint on nucleon FSI via a measurement of $\delta \alpha_T$. To evaluate this, Fig.~\ref{protonFSI_postfit} shows only the proton FSI strength uncertainty as a function of the number of POT following fits using either $\delta \alpha_T$ or $\delta p_T$ alongside $E_{vis}$. The extracted uncertainty on the nucleon FSI parameter is of the order of few percent at low statistics and can reach 1\% with Hyper-K-era statistics. As can be seen, $\delta p_T$ is only slightly less sensitive to FSI than $\delta\alpha_T$, but such sensitivity comes from the tail of the distribution with some degeneracy with the impact of non-QE components and SRCs. As such, the FSI sensitivity in $\delta p_T$ is more dependent on the shape uncertainties assumed for non-QE and less robust than the sensitivity from $\delta\alpha_T$. On the other hand we expect in a full multi-dimensional fit, using both $\delta p_T$ and $\delta \alpha_T$ to provide an even more robust constraint on FSI and to be able to cross-check the correctness of FSI simulations through the investigation of possible tensions between the two variables.


\begin{figure}[]
\begin{center}
\includegraphics[width=0.45\textwidth]{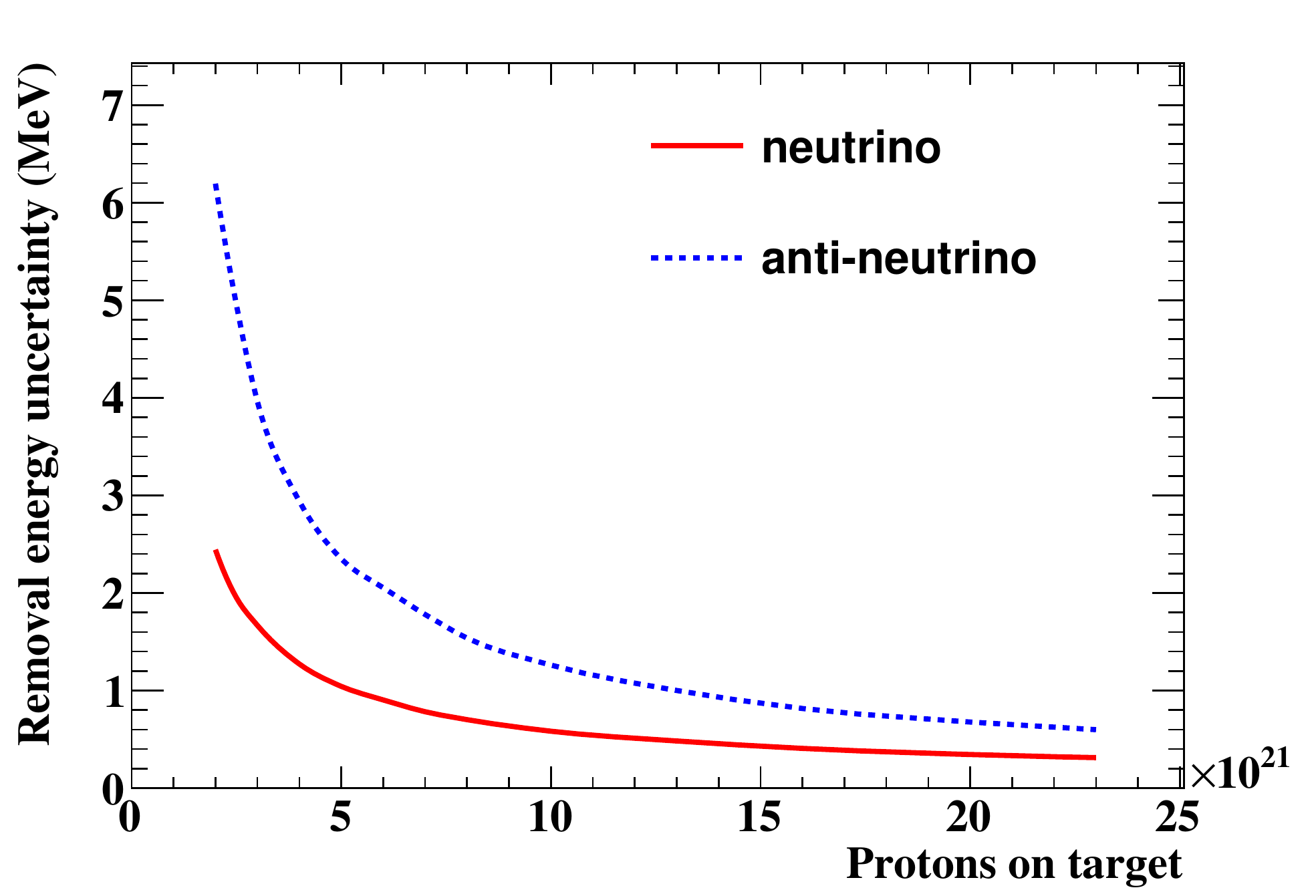}
\end{center}
\caption{The 1$\sigma$ sensitivity to the nuclear removal energy shift parameter as a function of POT for neutrino and anti-neutrino interactions when fitting the reconstructed CC0$\pi$ data binned in $\delta p_T$ and $E_{vis}$.}
\label{Eb_uncer}
\end{figure}

Fig.~\ref{Eb_uncer} shows the constraint on the removal energy parameter from a fit to $E_{vis}$ and $\delta p_T$ for neutrino and anti-neutrino interactions. In the neutrino case the removal energy shift can be measured at 2~MeV at relatively low statistics and better than 1~MeV with ultimate statistics. The corresponding anti-neutrino constraint is 3 to 4 times worse. As discussed in Appendix~\ref{app:pn}, further improved constraints can be obtained by exploiting the $p_N$ variable in place of $\delta p_T$, but in this case more longitudinal information is included and thus the constraint is more reliant on the neutrino flux shape prediction.

\begin{figure}[]
\begin{center}
\includegraphics[width=0.45\textwidth]{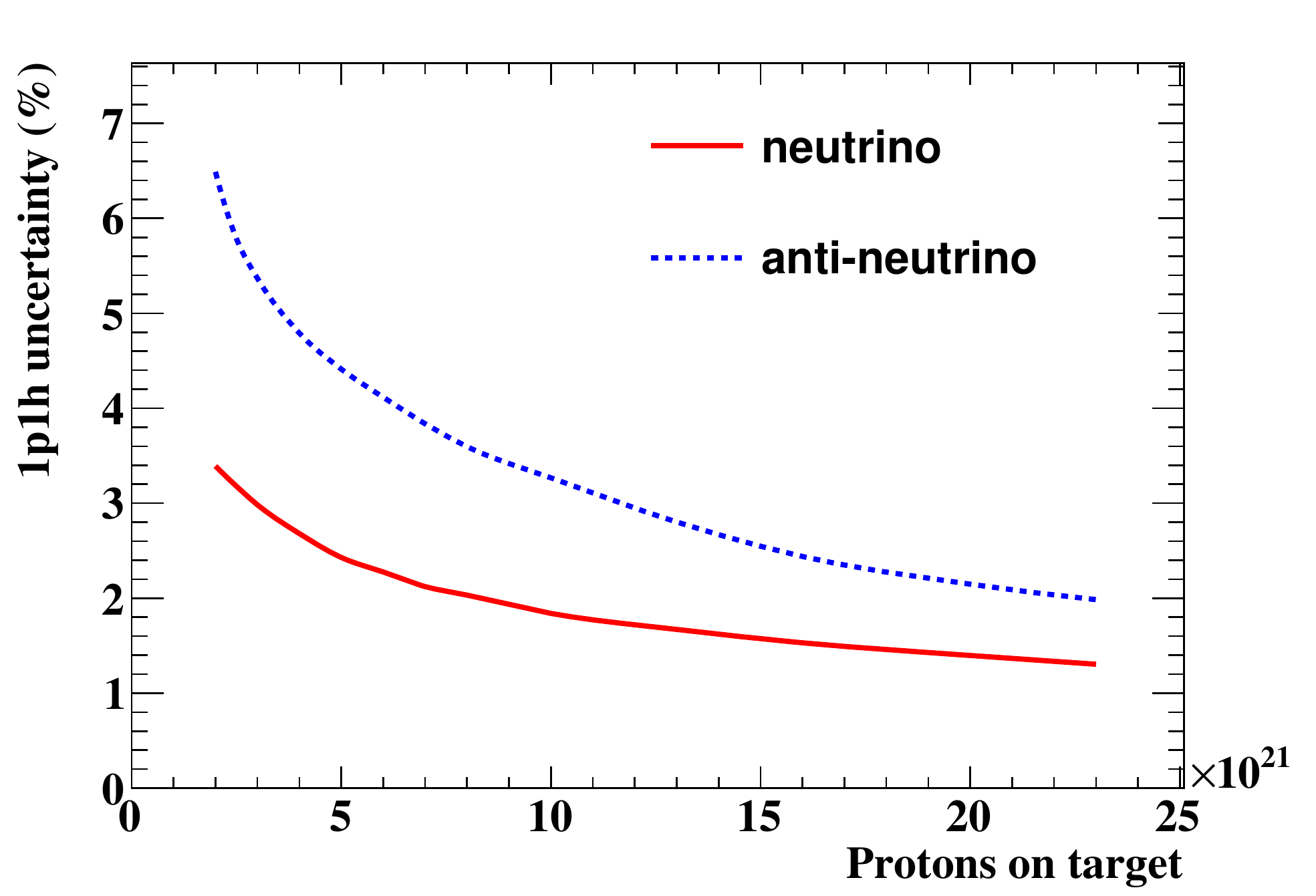}
\includegraphics[width=0.45\textwidth]{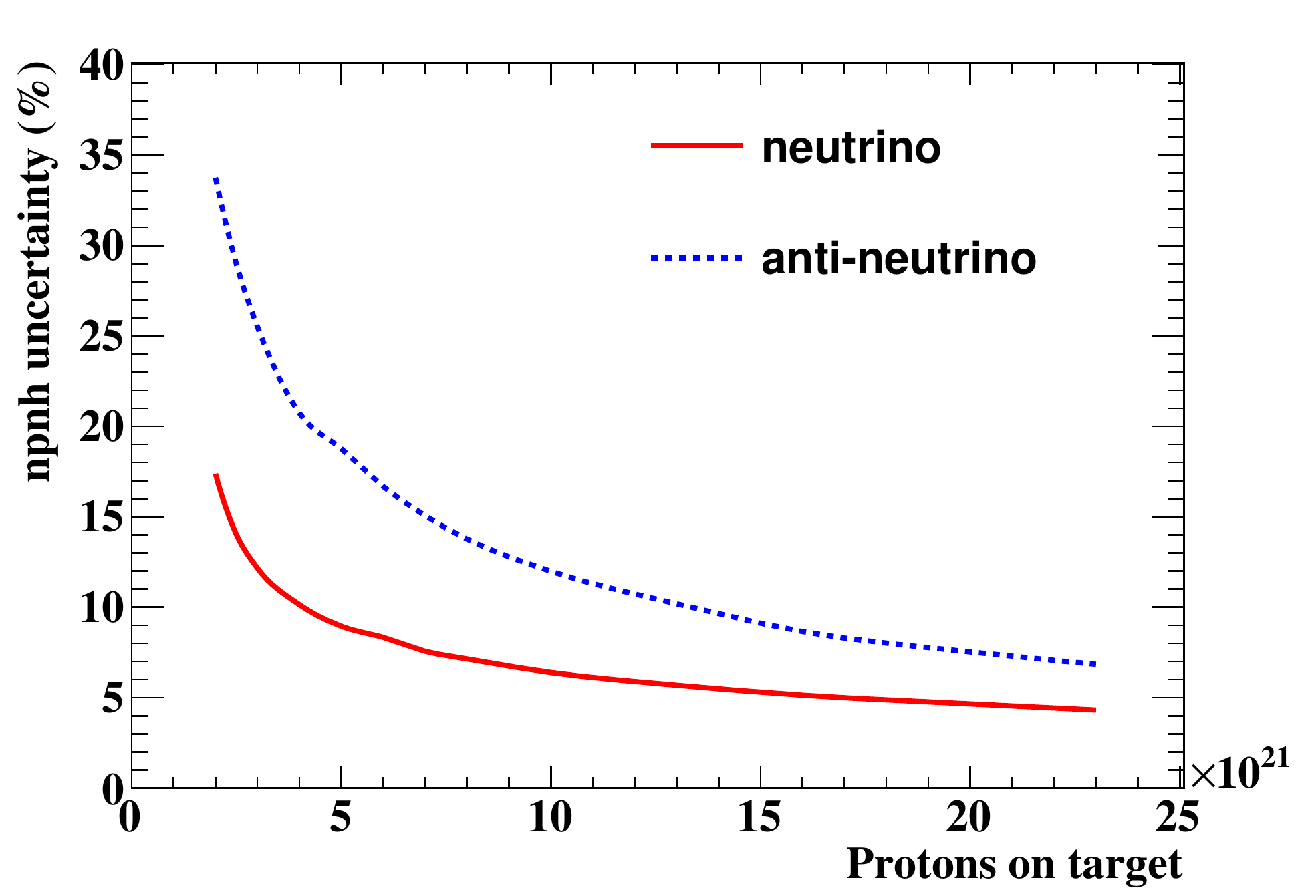}
\end{center}
\caption{The 1$\sigma$ sensitivity to the 1p1h (top) and npnh (bottom) cross-section normalisations as a function of POT for neutrino and anti-neutrino interactions when fitting the reconstructed CC0$\pi$ data binned in $\delta p_T$ and $E_{vis}$.}
\label{1p1h_npnh}
\end{figure}

As discussed in Sec~\ref{sec:variables}, non-QE and SRC interactions in the CC0$\pi$ channel can induce an important bias in the reconstruction of neutrino energy. Fig.~\ref{1p1h_npnh} shows the constraint on the total normalisation of 1p1h and npnh interactions. As previously noted, relatively advanced shape uncertainties are considered for these processes in the fit but for brevity we show only their combined effect on the total cross sections. The bulk of the $\delta p_T$ variable is a direct probe of the 1p1h interactions, allowing a constraint as good as $\sim1.5$\% ($\sim2$\%) in neutrino (anti-neutrino) interactions. The tail at high values of $\delta p_T$ is sensitive to the non-QE component, enabling a constraint better than 5\% (10\%) in neutrino (anti-neutrino) interactions. The npnh constraint is partially correlated with the constraint on the pion absorption FSI component and the background due to undetected pions in the final state. It should be noted that the effects of such components on the neutrino energy reconstruction are similar so induce similar bias in the neutrino oscillation parameters. Indeed, pion FSI and the 2p2h process including a $\Delta$ resonance excitation and a pion production followed by re-absorption, are separately modelled in NEUT but they are fundamentally very similar processes (i.e. they are partially irreducible backgrounds to each other). On the contrary, the background due to the presence of an undetected pion in the final state can be reduced, notably thanks to the lower threshold for pion reconstruction in the \sfgd.

The main effect limiting the precision of the extrapolation of the 1p1h normalisation from the near to the far detector is the degeneracy between the cross section and the neutrino flux. In Ref.~\cite{Munteanu:2019llq}, a strategy to select an hydrogen enhanced sample in anti-neutrino interactions has been proposed, in order to constrain the flux independently of uncertainties on nuclear effects. Here we consider the hydrogen sample together with the carbon sample in a joint fit to estimate quantitatively its impact on the flux constraints. The contribution of hydrogen interactions in the anti-neutrino sample is illustrated in Fig.~\ref{dpt_all_mode_anu} for the $\delta p_T$ distribution. 

\begin{figure}[h]
\begin{center}
\includegraphics[width=0.45\textwidth]{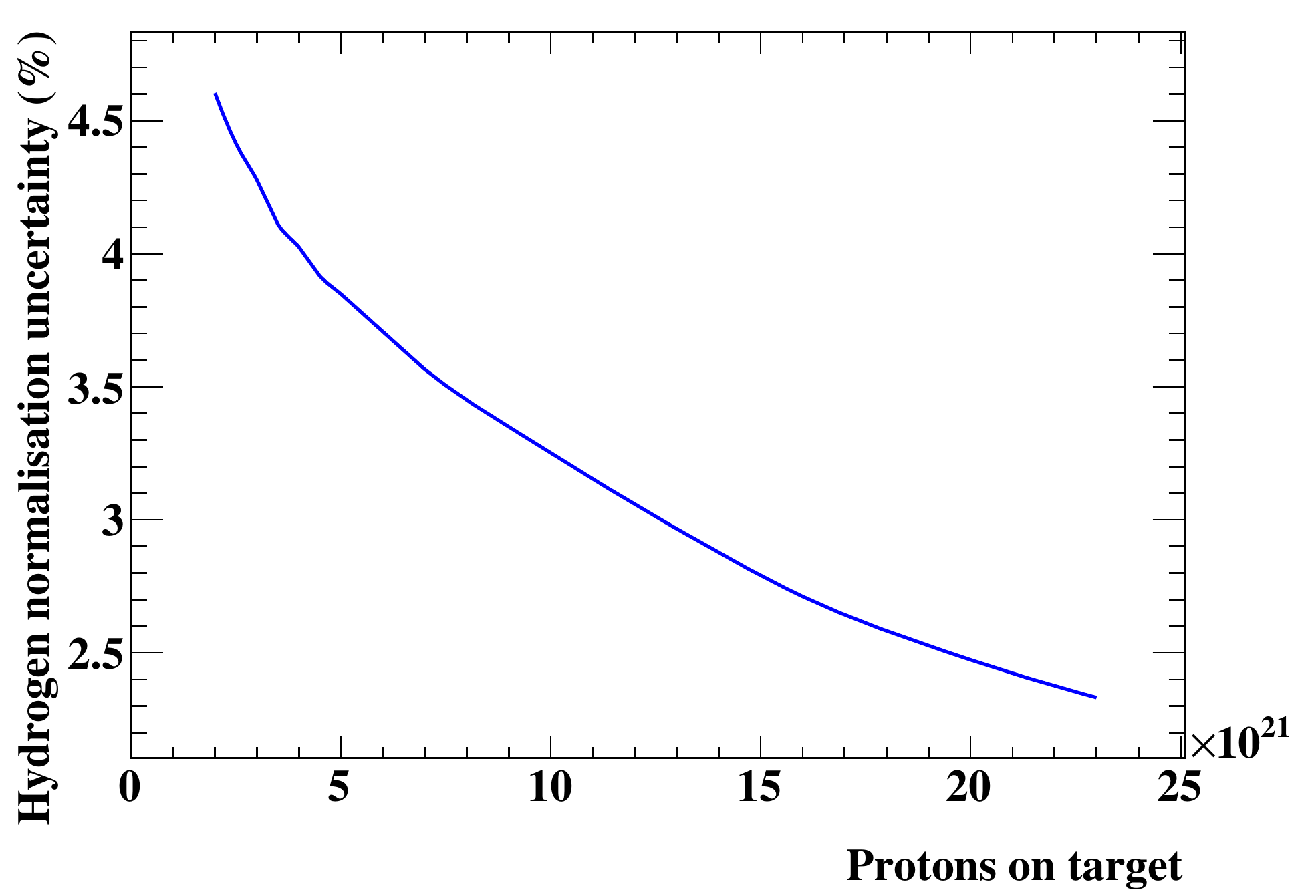}
\includegraphics[width=0.45\textwidth]{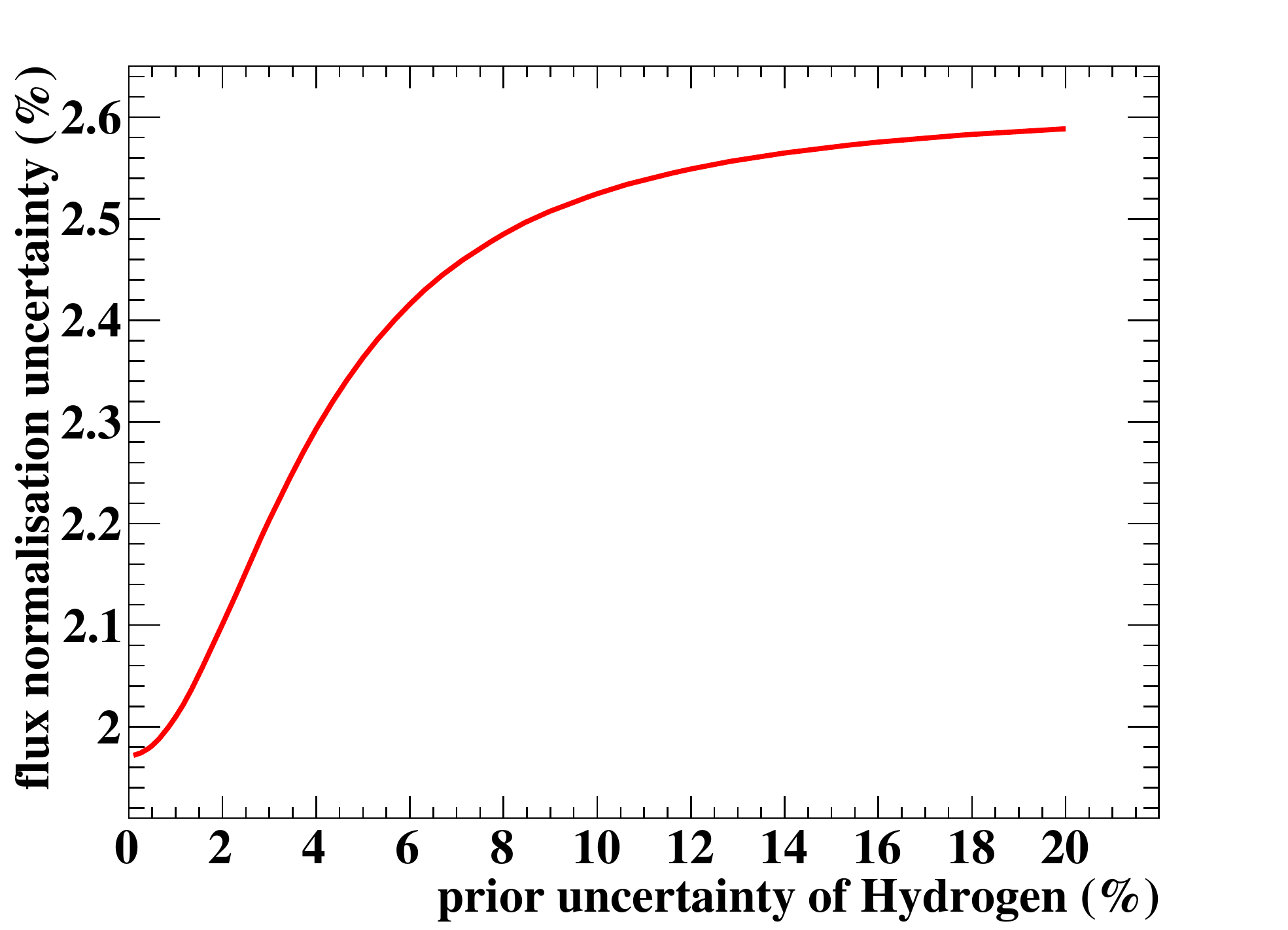}
\end{center}
\caption{The 1$\sigma$ sensitivity to the hydrogen normalisation parameter as a function of POT (top) and the 1$\sigma$ sensitivity to the flux normalisation as a function of the hydrogen parameter's prior uncertainty (bottom) for anti-neutrino interactions when fitting the reconstructed CC0$\pi$ data binned in $\delta p_T$ and $E_{vis}$ for $1\times 10^{22}$ POT.} 
\label{fig:hstudy}
\end{figure}

Fig.~\ref{fig:hstudy} shows the constraint on the total normalisation of the hydrogen sample, which is reduced by a factor of $\sim$2 relative to the prior uncertainty with $2\times 10^{22}$~POT worth of data, in addition to the constraint on the flux normalisation. The impact on the flux constraint depends on the prior uncertainties assumed on the anti-neutrino-nucleon cross section, notably due to the nucleon form-factor. It should be noted that this is the result of a complete fit including both anti-neutrino-nucleus and anti-neutrino-nucleon processes, thus showing the {\it incremental} contribution to the flux precision from the hydrogen enhanced sample, on top of the constraint obtained from the usual fit to carbon interactions. The contribution of the hydrogen-enhanced sample is sizeable, if the prior hydrogen normalisation uncertainty is lower than $\sim$10\%, improving the relative flux constraint by up to $\sim$20\%. Note that, whilst not quantified here, the power of the hydrogen-enhanced sample also improves knowledge of the flux shape and that better separation between flux and form-factor constraints could be achieved by adding the reconstructed four-momentum ($Q^2$) from the lepton kinematics as a fit variable. This would allow a separation of the low $Q^2$ region, in which the form-factor is relatively well understood and so the flux can be constrained, from the high $Q^2$ region, where the prior flux constraint can instead be leveraged to probe the form-factor details.

\begin{table*}[htbp]
\centering
\begin{tabular}{|l|c|c|c|}
\hline
$1 \times 10^{22}$POT & {$\delta p_T$;$E_{vis}$} & {$\delta\alpha_T$;$E_{vis}$} & {$p_N$;$E_{vis}$}  \\
\hline
\rowcolor{LightCyan}
1p1h ($\nu$) & 1.9\% & 1.8\% & 1.5\% \\
\hline
1p1h ($\nub$) & 3.3\% & 3.9\% & 2.6\% \\
\hline
\rowcolor{LightCyan}
npnh ($\nu$) & 6.5\% & 13\% & 5.3\% \\
\hline
npnh ($\nub$) & 12\% & 17\% & 11\% \\
\hline
\rowcolor{LightCyan}
$E_{rmv}$ ($\nu$)  & 0.55 MeV & 0.38 MeV & 0.53 MeV \\
\hline
$E_{rmv}$ ($\nub$) & 1.3 MeV & 1.0 MeV & 1.3 MeV \\
\hline
\rowcolor{LightCyan}
Pion FSI ($\nu$) & 6.6\% & 14\% & 4.8\% \\
\hline
Pion FSI ($\nub$) & 34\% & 35\% & 30\% \\
\hline
\rowcolor{LightCyan}
Undetected pions ($\nu$) & 9.7\% & 14\% & 8.2\% \\
\hline
Undetected pions ($\nub$) & 37\% & 36\% & 31\% \\
\hline
\rowcolor{LightCyan}
Nucleon FSI ($\nu$) & 1.1\% & 0.76\%  & 0.98\% \\
\hline
Nucleon FSI ($\nub$) & 2.3\% & 1.9\% & 2.4\%\\
\hline
\rowcolor{LightCyan}
Flux ($\nu$) & 1.8\% & 1.9\% & 1.6\%  \\
\hline
Flux ($\nub$) & 2.4\% & 2.3\% & 2.2\%\\
\hline
\rowcolor{LightCyan}
Total ($\nu$) & 1.8\% & 2.1\% & 1.6\% \\
\hline
Total ($\nub$) & 2.7\% & 2.7\% & 2.5\%\\
\hline
\rowcolor{LightCyan}
Hydrogen ($\nub$) & 3.3\% & 4.0\% & 2.9\% \\
\hline
\end{tabular}
\hspace{10mm}
\begin{tabular}{|l|c|c|c|}
\hline
$2 \times 10^{22}$POT & {$\delta p_T$;$E_{vis}$} & {$\delta\alpha_T$;$E_{vis}$} & {$p_N$;$E_{vis}$}  \\
\hline
\rowcolor{LightCyan}
1p1h ($\nu$) & 1.4\% & 1.2\% & 1.1\%\\
\hline
1p1h ($\nub$) & 2.2\% & 2.5\% & 1.7\%\\
\hline
\rowcolor{LightCyan}
npnh ($\nu$) & 4.7\% & 8.5\% & 3.6\% \\
\hline
npnh ($\nub$) & 7.5\% & 10.0\% & 7.1\% \\
\hline
\rowcolor{LightCyan}
$E_{rmv}$ ($\nu$) &  0.32 MeV & 0.22 MeV &  0.3 MeV \\
\hline
$E_{rmv}$ ($\nub$) & 0.68 MeV & 0.57 MeV & 0.68 MeV \\
\hline
\rowcolor{LightCyan}
Pion FSI ($\nu$) & 4.2\% & 8.3\% & 2.8\%\\
\hline
Pion FSI ($\nub$) & 19\% & 17\% & 16\%  \\
\hline
\rowcolor{LightCyan}
Undetected pions ($\nu$) & 6.6\% & 7.9\% & 5.4\%\\
\hline
Undetected pions ($\nub$) & 21\% & 19\% & 18\% \\
\hline
\rowcolor{LightCyan}
Nucleon FSI ($\nu$) & 0.62\% & 0.43\% & 0.58\%  \\
\hline
Nucleon FSI ($\nub$) & 1.2\% & 0.98\% & 1.2\%  \\
\hline
\rowcolor{LightCyan}
Flux ($\nu$) & 1.4\% & 1.4\% & 1.2\%  \\
\hline
Flux ($\nub$) & 1.7\% & 1.6\% & 1.6\% \\
\hline
\rowcolor{LightCyan}
Total ($\nu$)  & 1.4\% & 1.5\% & 1.2\%  \\
\hline
Total ($\nub$) & 1.8\% & 1.7\% & 1.7\% \\
\hline
\rowcolor{LightCyan}
Hydrogen ($\nub$) & 2.5\% & 3.1\% & 2.1\% \\
\hline
\end{tabular}
\caption{\label{tab:para_uncert1} Expected 1$\sigma$ uncertainties on key cross-section parameters at $1 \times 10^{22}$ POT (left) and $2 \times 10^{22}$ POT (right) for neutrino and anti-neutrino interactions with different fit variables.} 
\end{table*}





\section{Conclusion}
\label{sec:conclusion}
It has been demonstrated that the upgraded ND280 detector will allow {\it increasing} sensitivity to nuclear-model uncertainties, enabled by the use of observables formed from both lepton and nucleon kinematics, which can complement the usual T2K analysis of only muon kinematics. Importantly, the use of nucleon kinematics also allows less room for incorrect models to describe ND280 Upgrade data and, as such, facilitates more robust constraints.

The inclusion of nucleon kinematics into the analysis brings new systematic uncertainties, notably detector systematics but also new nuclear-model systematics related with nucleon FSI. From simulated studies, benchmarked by test beam data of prototypes and by long-term data-taking experience with ND280, we expect the systematic uncertainties related with detector modelling to be well under control. In this paper we have shown quantitatively, for the first time, that nucleon FSI can also be very well constrained thanks to the use $\delta\alpha_T$. The precision of the constraint is enabled by the low proton (and neutron) tracking threshold in ND280 Upgrade and the absence of degeneracy (correlation) with other nuclear-model uncertainties in $\delta\alpha_T$.

The use of an improved estimator of neutrino energy, based on the sum of muon energy and nucleon kinetic energy, has been investigated and shows interesting sensitivity to nuclear removal energy shifts. Furthermore, the QE and non-QE components of the cross section have been shown to be well separated by using fits to $\delta p_T$. Contrary to an inclusive analysis, these two components can be measured with small degeneracy/correlation, thus reducing ambiguities in the propagation of constraints to the far detector. Finally, the reduction of the flux uncertainty from a hydrogen-enhanced sample, as suggested in Ref.~\cite{Munteanu:2019llq}, has been quantified for the first time. The relative improvement on the flux normalisation uncertainty can be up to 20\%, depending on the prior uncertainty on the nucleon form factors. 

The sensitivity studies presented here are based on the simulation of the detector performances expected with ND280 Upgrade and includes a relatively sophisticated, but still incomplete, set of nuclear uncertainties. Conservative assumptions have been taken when possible, e.g., considering mostly the kinematics projected in the transverse plane to minimise the dependence on flux modelling, leaving parameters unconstrained, and fitting the neutrino and anti-neutrino samples separately, thus not including the correlation between neutrino and anti-neutrino cross-section and flux uncertainties. Overall the results demonstrate the interesting potential model constraints enabled by exploiting proton and neutron kinematics measured at the near detector for future oscillation analyses using ND280 Upgrade.

Whilst it is shown that most parameter constraints plateau at an acceptable level for Hyper-K era statistics, some show scope for improvement even beyond this limit. This is particularly true for those relating to the flux constraint from the hydrogen enhanced region of $\delta p_T$. At very high statistics improved (and more robust) sensitivities may also be gained by using finer binning or adding additional dimensions to the fit. It should be noted that this could be realised by either additional beam exposure for ND280 Upgrade and/or future further upgrades to increase the target mass (e.g. using the methods suggested in Refs.~\cite{Berns:2020ehg,Boyarintsev:2021uyw}).

\vspace{-4mm}

\section*{Acknowledgements}
\vspace{-2mm}

This work was initiated in the framework of the T2K Near Detector upgrade project, fruitful discussions in this context with our colleagues are gratefully acknowledged. This project has received funding from the European Union’s Horizon 2020 research and innovation programme under the Marie Skłodowska-Curie grant agreement No 754496. It also received support from the Russian Science Foundation grant \#19-12-00325 and  the support in the framework of the State project ``Science'' by the Ministry of Science and Higher Education of the Russian Federation under the contract 075-15-2020-778. This work was supported by P2IO LabEx (ANR-10-LABX-0038 – Project ``BSMNu'') in the framework ``Investissements d’Avenir'' (ANR-11-IDEX-0003-01 ) managed by the Agence Nationale de la Recherche (ANR), France, in addition to Agence Nationale de la Recherche (ANR), France, grant no. ANR-19-CE31-0001 - project "SUNCORE". We acknowledge the support of CNRS/IN2P3 and CEA.


\appendix

\section{Fitting with reconstructed neutron momentum}
\label{app:pn}

As discussed in Sec.~\ref{sec:variables}, the reconstructed nucleon momentum ($p_N$)~\cite{Furmanski:2016wqo} can be used as an alternative to the transverse momentum imbalance ($\delta p_T$)~\cite{Lu:2015tcr} as an input variable in the fit. The former contains more information than that latter, incorporating an inferred longitudinal imbalance, in addition to the transverse imbalance, and so better sensitivity is in principle expected. However, this better sensitivity accompanies the potential need for additional systematic uncertainties. For example, whilst the shape of $\delta p_T$ is largely independent from neutrino energy and nucleon-level physics~\cite{Lu:2015tcr}, this will not be the case for $p_N$. We therefore choose the more conservative and simple approach of quoting the primary sensitivities using a fit in $\delta p_T$ but discuss here what would be obtained using $p_N$ instead.

\begin{figure*}
\begin{center}
\includegraphics[width=0.45\textwidth]{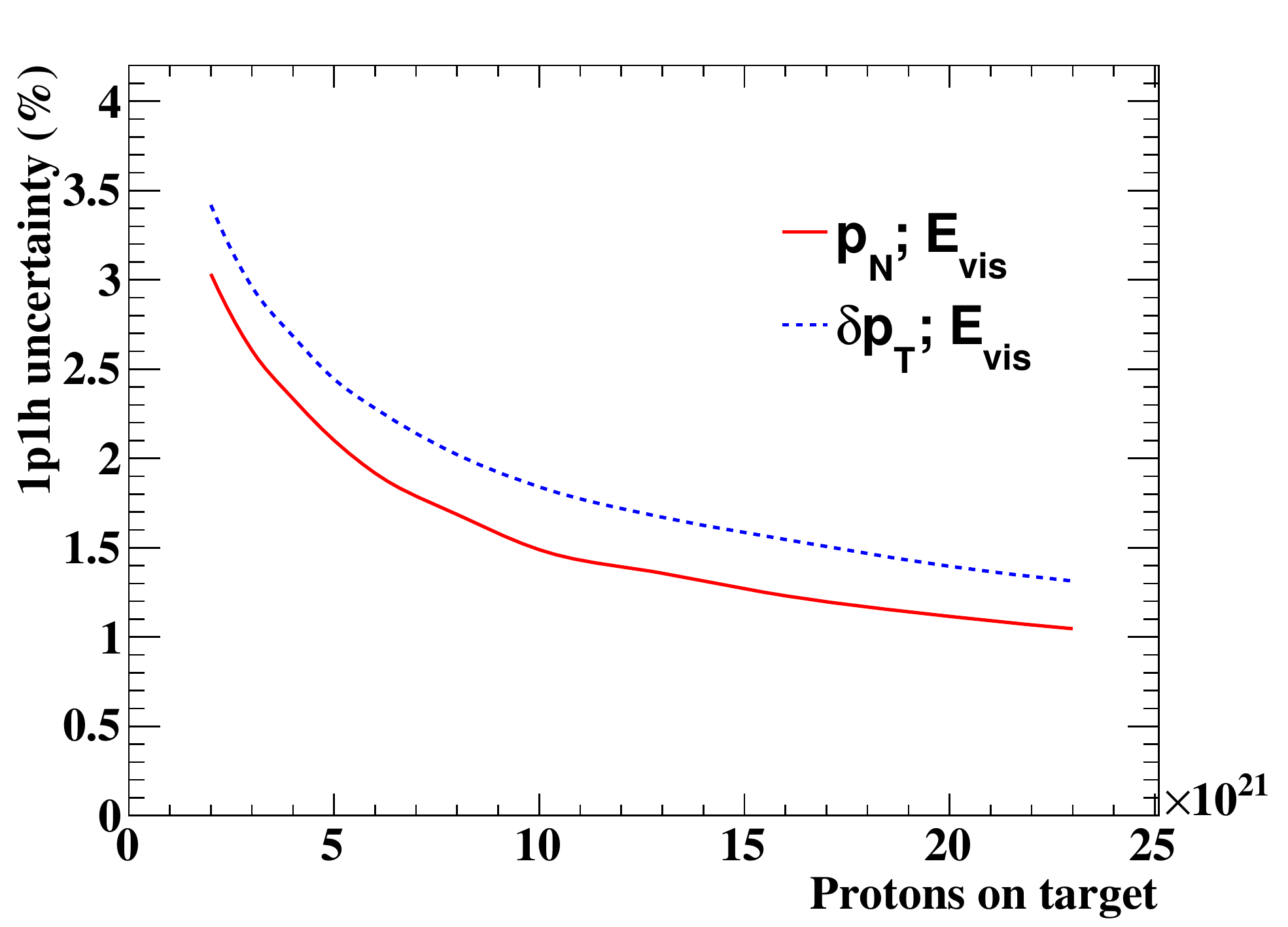}
\includegraphics[width=0.45\textwidth]{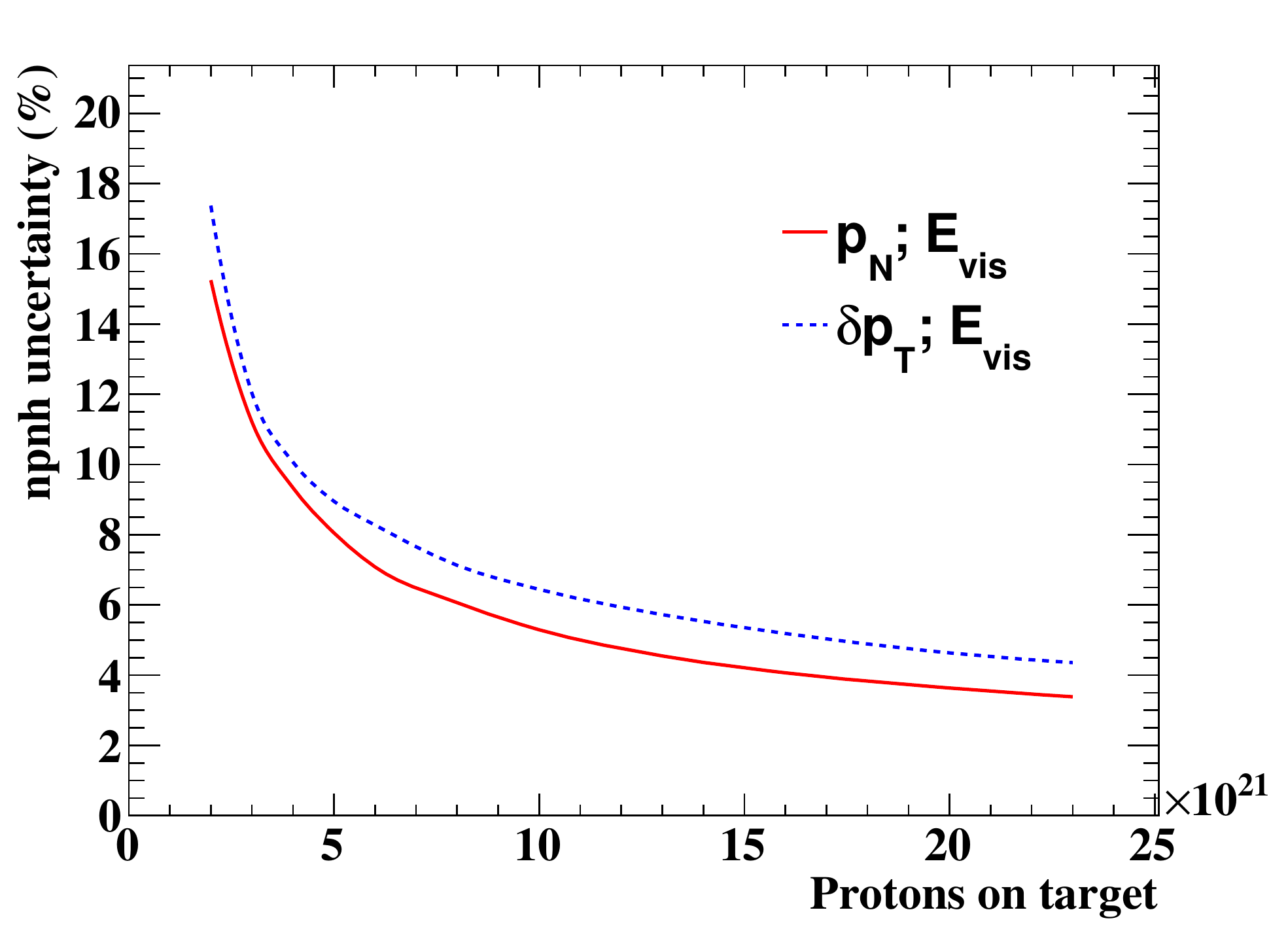}
\includegraphics[width=0.45\textwidth]{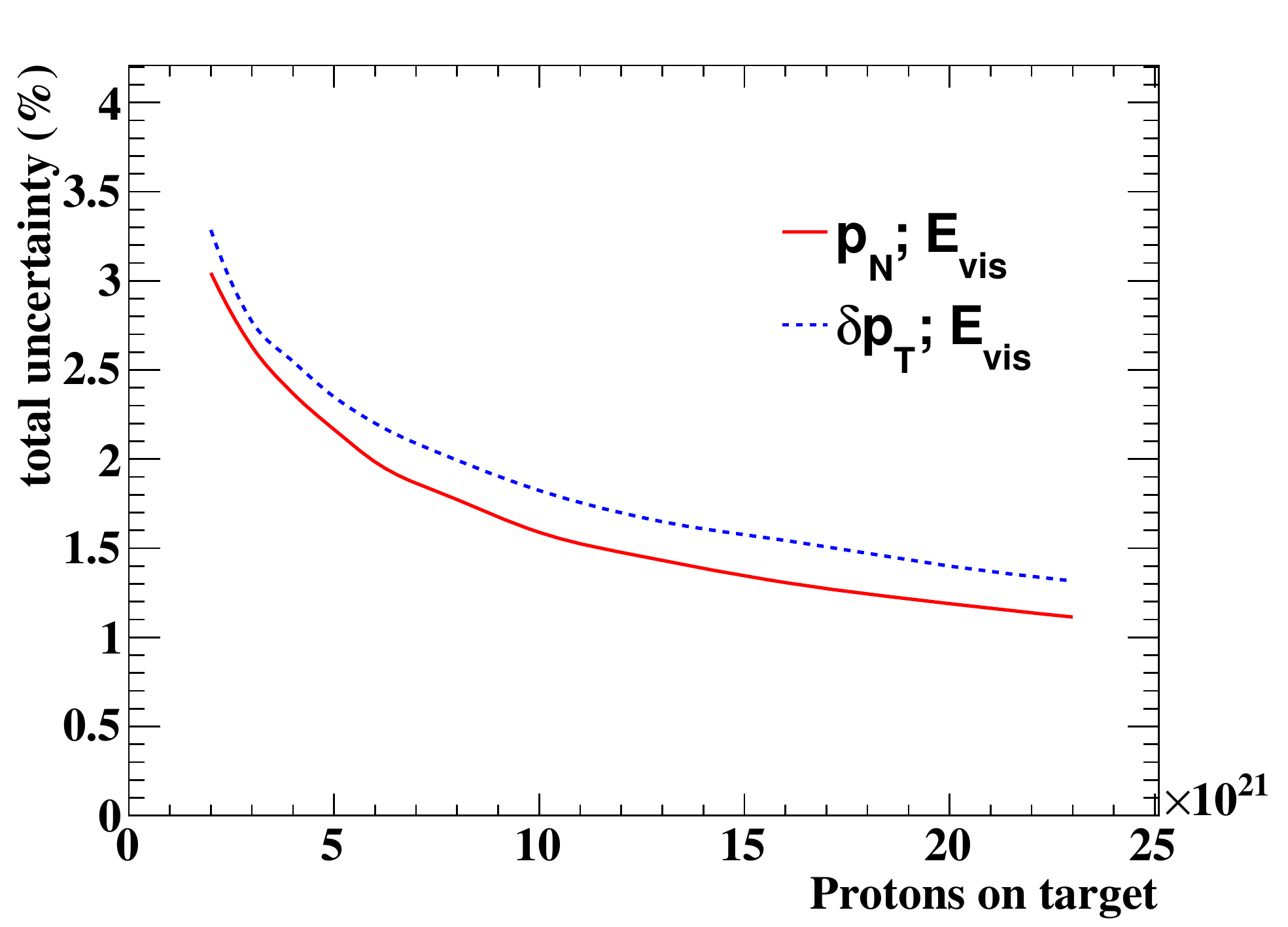}
\end{center}
\caption{ The 1$\sigma$ sensitivity to the 1p1h (upper), npnh (middle) and total (lower) cross-section normalisations as a function of POT for neutrino and anti-neutrino interactions when fitting the reconstructed CC0$\pi$ data binned in $E_{vis}$ and either $\delta p_T$ or $p_N$.} 
\label{fig:pn_dpt}
\end{figure*}

Generally it is found that using $p_N$ in place of $\delta p_T$ has a relatively modest impact on the extracted sensitivities shown in Sec.~\ref{sec:results}, although a notable improvement was found for some parameters. These improvements are shown in Fig.~\ref{fig:pn_dpt}, demonstrating how $p_N$ offers better ability to distinguish npnh, 1p1h and pion production processes compared to $\delta p_T$. This behaviour has previously been noted in e.g. Ref.~\cite{Dolan:2018zye}.


\vspace{-2mm}
\section{Histograms input to the fit}
\label{app:allInputs}
\begin{figure*}
\begin{center}
\includegraphics[width=0.32\textwidth]{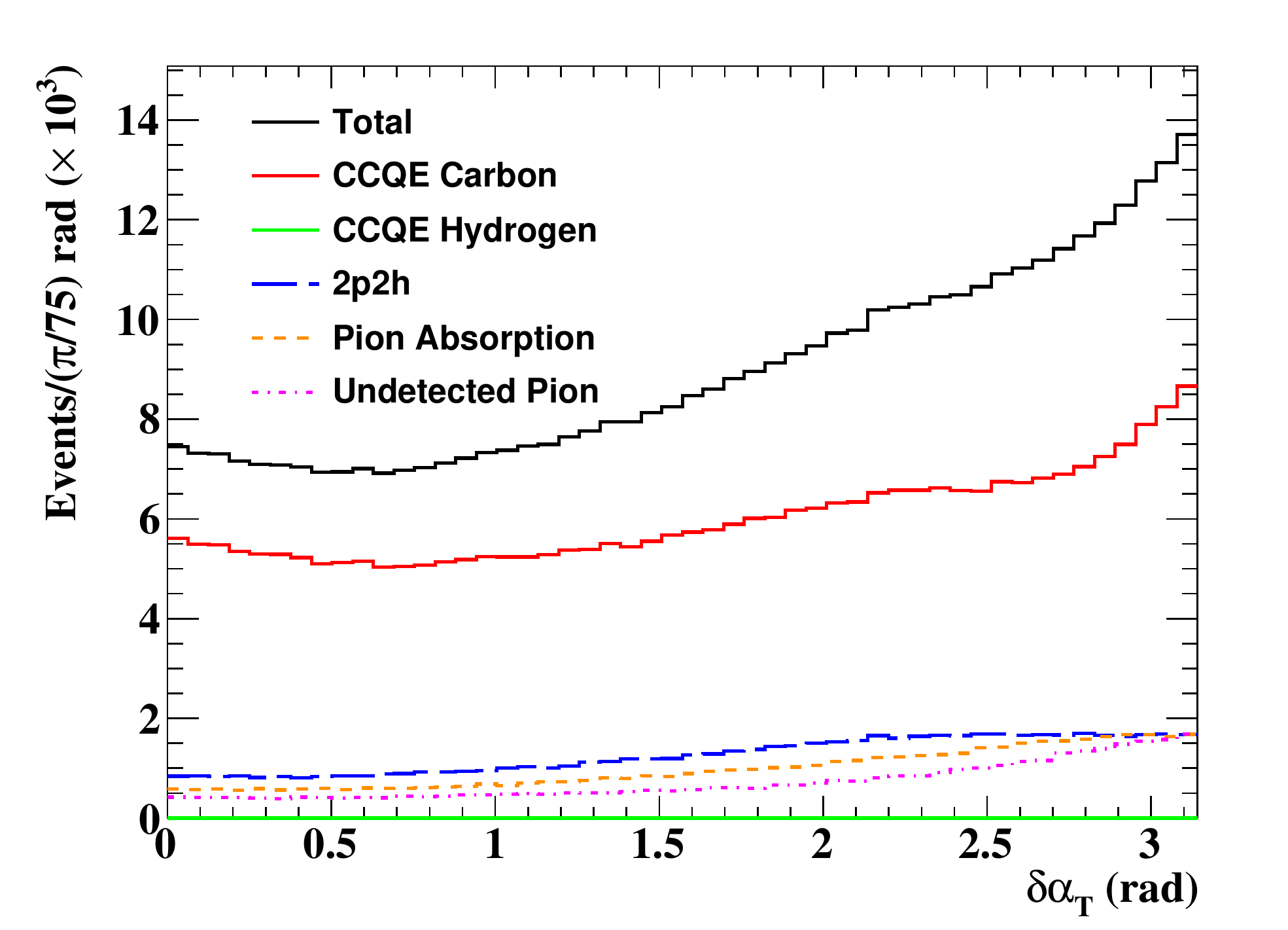}
\includegraphics[width=0.32\textwidth]{plots/input_mode/dpt_nu_POT1E22_bin30MeV.pdf}
\includegraphics[width=0.32\textwidth]{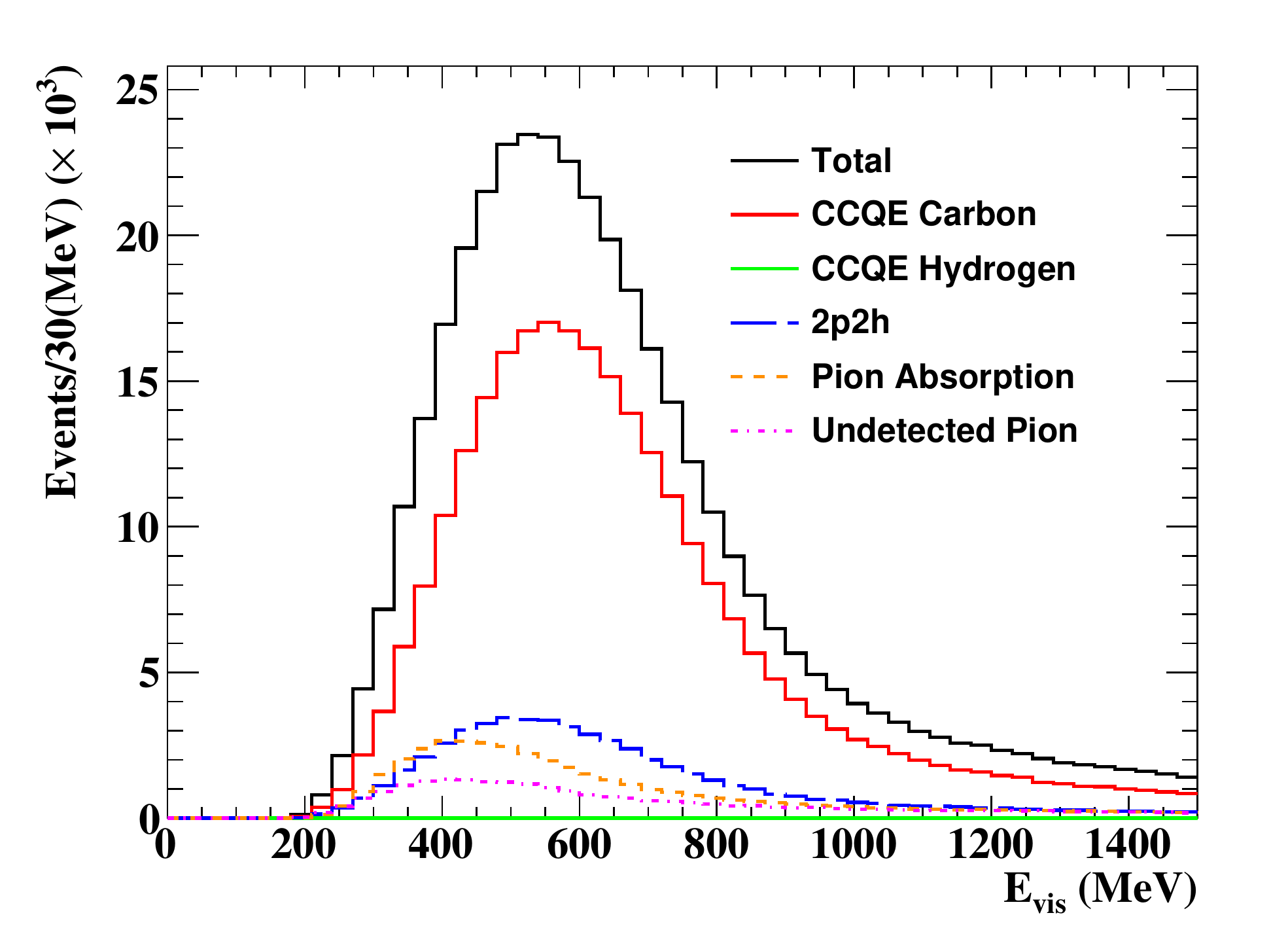}
\includegraphics[width=0.32\textwidth]{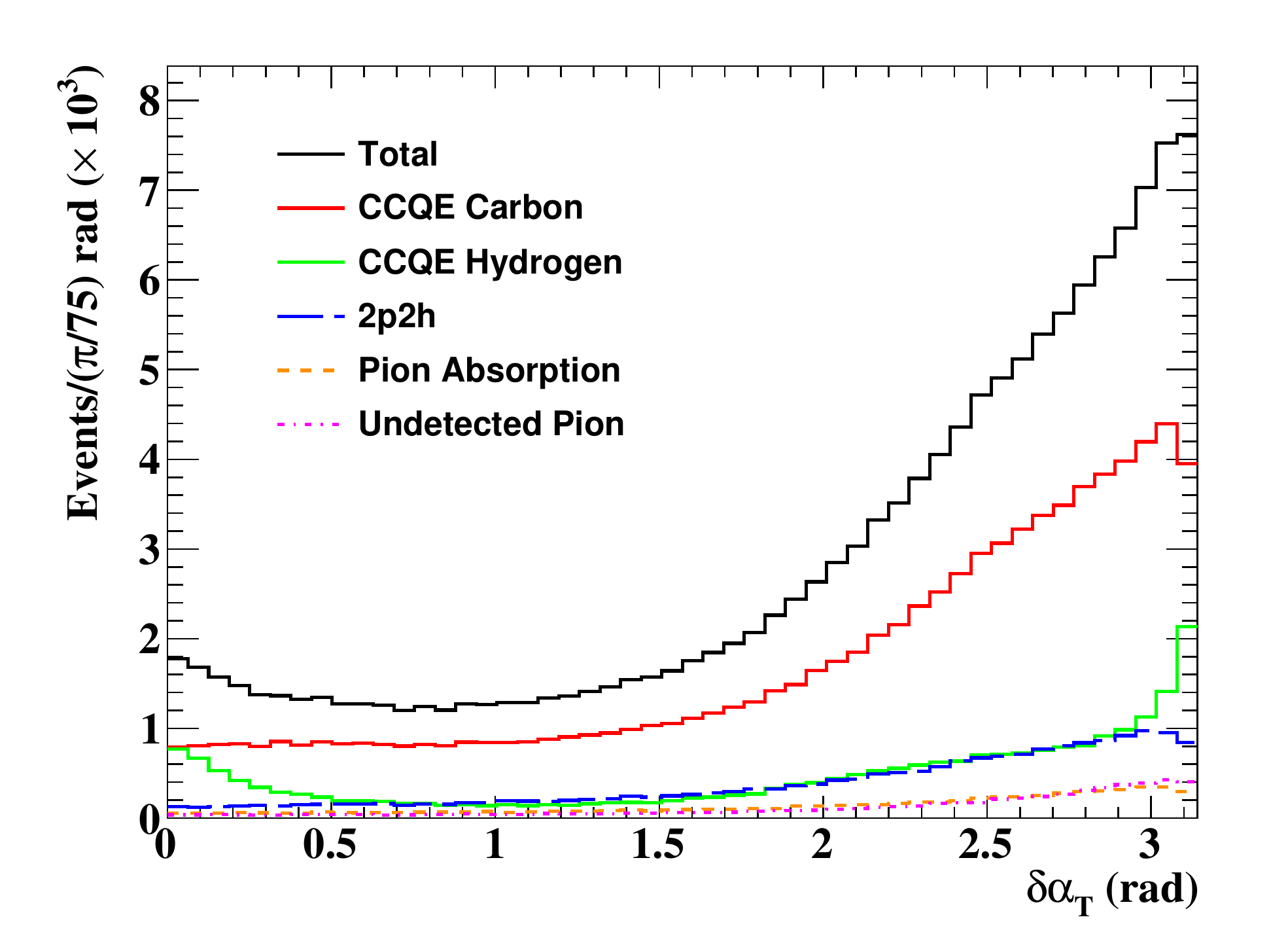}
\includegraphics[width=0.32\textwidth]{plots/input_mode/dpt_anu_POT1E22_bin30MeV.pdf}
\includegraphics[width=0.32\textwidth]{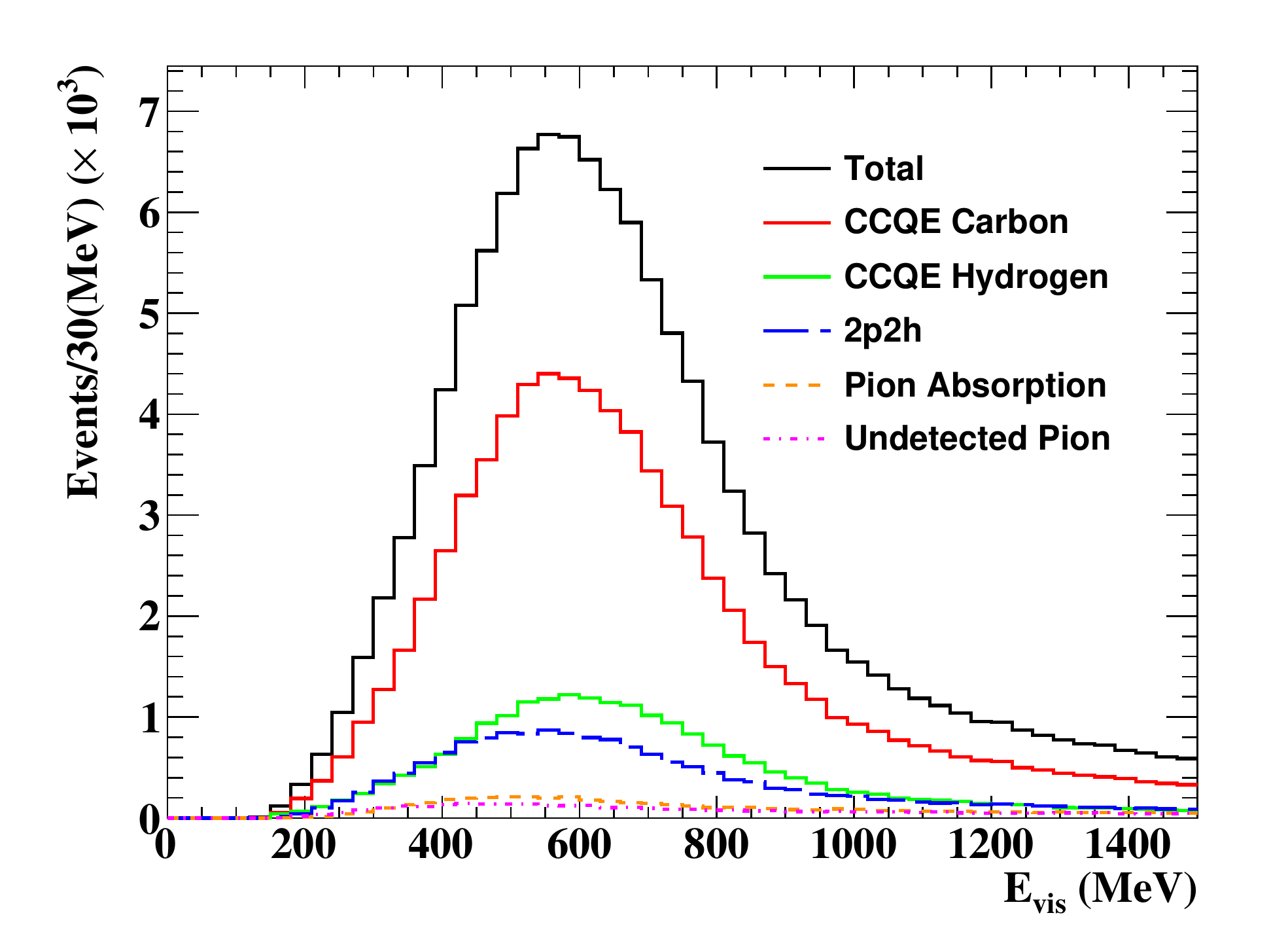}
\end{center}
\caption{One dimensional projections of the two-dimensional input histograms used to determine the sensitivities, shown here for $1 \times 10^{22}$ POT for neutrino (top) and anti-neutrino (bottom) interactions. The binning shown is that which is used in the fits. }
\label{fig:allInputs}
\end{figure*}

The full set of input histograms used in the fit projected onto single dimensions for neutrino and anti-neutrino interactions are shown in Fig.~\ref{fig:allInputs}. The binning matches the one used in the fits. The histograms are shown in reconstructed kinematic quantities following the CC0$\pi$ event selection and are broken down by interaction mode.

\bibliography{biblio}

 \newcommand{\noop}[1]{}
\begin{thebibliography}{37}%
\makeatletter
\providecommand \@ifxundefined [1]{%
 \@ifx{#1\undefined}
}%
\providecommand \@ifnum [1]{%
 \ifnum #1\expandafter \@firstoftwo
 \else \expandafter \@secondoftwo
 \fi
}%
\providecommand \@ifx [1]{%
 \ifx #1\expandafter \@firstoftwo
 \else \expandafter \@secondoftwo
 \fi
}%
\providecommand \natexlab [1]{#1}%
\providecommand \enquote  [1]{``#1''}%
\providecommand \bibnamefont  [1]{#1}%
\providecommand \bibfnamefont [1]{#1}%
\providecommand \citenamefont [1]{#1}%
\providecommand \href@noop [0]{\@secondoftwo}%
\providecommand \href [0]{\begingroup \@sanitize@url \@href}%
\providecommand \@href[1]{\@@startlink{#1}\@@href}%
\providecommand \@@href[1]{\endgroup#1\@@endlink}%
\providecommand \@sanitize@url [0]{\catcode `\\12\catcode `\$12\catcode
  `\&12\catcode `\#12\catcode `\^12\catcode `\_12\catcode `\%12\relax}%
\providecommand \@@startlink[1]{}%
\providecommand \@@endlink[0]{}%
\providecommand \url  [0]{\begingroup\@sanitize@url \@url }%
\providecommand \@url [1]{\endgroup\@href {#1}{\urlprefix }}%
\providecommand \urlprefix  [0]{URL }%
\providecommand \Eprint [0]{\href }%
\providecommand \doibase [0]{http://dx.doi.org/}%
\providecommand \selectlanguage [0]{\@gobble}%
\providecommand \bibinfo  [0]{\@secondoftwo}%
\providecommand \bibfield  [0]{\@secondoftwo}%
\providecommand \translation [1]{[#1]}%
\providecommand \BibitemOpen [0]{}%
\providecommand \bibitemStop [0]{}%
\providecommand \bibitemNoStop [0]{.\EOS\space}%
\providecommand \EOS [0]{\spacefactor3000\relax}%
\providecommand \BibitemShut  [1]{\csname bibitem#1\endcsname}%
\let\auto@bib@innerbib\@empty
\bibitem [{\citenamefont {Alvarez-Ruso}\ \emph {et~al.}(2018)\citenamefont
  {Alvarez-Ruso} \emph {et~al.}}]{Alvarez-Ruso:2017oui}%
  \BibitemOpen
  \bibfield  {author} {\bibinfo {author} {\bibfnamefont {L.}~\bibnamefont
  {Alvarez-Ruso}} \emph {et~al.},\ }\href {\doibase 10.1016/j.ppnp.2018.01.006}
  {\bibfield  {journal} {\bibinfo  {journal} {Prog. Part. Nucl. Phys.}\
  }\textbf {\bibinfo {volume} {100}},\ \bibinfo {pages} {1} (\bibinfo {year}
  {2018})},\ \Eprint {http://arxiv.org/abs/1706.03621} {arXiv:1706.03621
  [hep-ph]} \BibitemShut {NoStop}%
\bibitem [{\citenamefont {Abe}\ \emph {et~al.}(2011{\natexlab{a}})\citenamefont
  {Abe} \emph {et~al.}}]{T2K:2011qtm}%
  \BibitemOpen
  \bibfield  {author} {\bibinfo {author} {\bibfnamefont {K.}~\bibnamefont
  {Abe}} \emph {et~al.} (\bibinfo {collaboration} {T2K}),\ }\href {\doibase
  10.1016/j.nima.2011.06.067} {\bibfield  {journal} {\bibinfo  {journal} {Nucl.
  Instrum. Meth. A}\ }\textbf {\bibinfo {volume} {659}},\ \bibinfo {pages}
  {106} (\bibinfo {year} {2011}{\natexlab{a}})},\ \Eprint
  {http://arxiv.org/abs/1106.1238} {arXiv:1106.1238 [physics.ins-det]}
  \BibitemShut {NoStop}%
\bibitem [{\citenamefont {Abe}\ \emph {et~al.}(2018{\natexlab{a}})\citenamefont
  {Abe} \emph {et~al.}}]{Hyper-Kamiokande:2018ofw}%
  \BibitemOpen
  \bibfield  {author} {\bibinfo {author} {\bibfnamefont {K.}~\bibnamefont
  {Abe}} \emph {et~al.} (\bibinfo {collaboration} {Hyper-Kamiokande}),\
  }\href@noop {} {\  (\bibinfo {year} {2018}{\natexlab{a}})},\ \Eprint
  {http://arxiv.org/abs/1805.04163} {arXiv:1805.04163 [physics.ins-det]}
  \BibitemShut {NoStop}%
\bibitem [{\citenamefont {Amaudruz}\ \emph {et~al.}(2012)\citenamefont
  {Amaudruz} \emph {et~al.}}]{T2KND280FGD:2012umz}%
  \BibitemOpen
  \bibfield  {author} {\bibinfo {author} {\bibfnamefont {P.~A.}\ \bibnamefont
  {Amaudruz}} \emph {et~al.} (\bibinfo {collaboration} {T2K ND280 FGD}),\
  }\href {\doibase 10.1016/j.nima.2012.08.020} {\bibfield  {journal} {\bibinfo
  {journal} {Nucl. Instrum. Meth. A}\ }\textbf {\bibinfo {volume} {696}},\
  \bibinfo {pages} {1} (\bibinfo {year} {2012})},\ \Eprint
  {http://arxiv.org/abs/1204.3666} {arXiv:1204.3666 [physics.ins-det]}
  \BibitemShut {NoStop}%
\bibitem [{\citenamefont {Abgrall}\ \emph {et~al.}(2011)\citenamefont {Abgrall}
  \emph {et~al.}}]{T2KND280TPC:2010nnd}%
  \BibitemOpen
  \bibfield  {author} {\bibinfo {author} {\bibfnamefont {N.}~\bibnamefont
  {Abgrall}} \emph {et~al.} (\bibinfo {collaboration} {T2K ND280 TPC}),\ }\href
  {\doibase 10.1016/j.nima.2011.02.036} {\bibfield  {journal} {\bibinfo
  {journal} {Nucl. Instrum. Meth. A}\ }\textbf {\bibinfo {volume} {637}},\
  \bibinfo {pages} {25} (\bibinfo {year} {2011})},\ \Eprint
  {http://arxiv.org/abs/1012.0865} {arXiv:1012.0865 [physics.ins-det]}
  \BibitemShut {NoStop}%
\bibitem [{\citenamefont {Abe}\ \emph {et~al.}(2019)\citenamefont {Abe} \emph
  {et~al.}}]{T2K:2019bbb}%
  \BibitemOpen
  \bibfield  {author} {\bibinfo {author} {\bibfnamefont {K.}~\bibnamefont
  {Abe}} \emph {et~al.} (\bibinfo {collaboration} {T2K}),\ }\href@noop {} {\
  (\bibinfo {year} {2019})},\ \Eprint {http://arxiv.org/abs/1901.03750}
  {arXiv:1901.03750 [physics.ins-det]} \BibitemShut {NoStop}%
\bibitem [{\citenamefont {Atti\'e}\ \emph {et~al.}(2020)\citenamefont {Atti\'e}
  \emph {et~al.}}]{Attie:2019hua}%
  \BibitemOpen
  \bibfield  {author} {\bibinfo {author} {\bibfnamefont {D.}~\bibnamefont
  {Atti\'e}} \emph {et~al.},\ }\href {\doibase 10.1016/j.nima.2019.163286}
  {\bibfield  {journal} {\bibinfo  {journal} {Nucl. Instrum. Meth. A}\ }\textbf
  {\bibinfo {volume} {957}},\ \bibinfo {pages} {163286} (\bibinfo {year}
  {2020})},\ \Eprint {http://arxiv.org/abs/1907.07060} {arXiv:1907.07060
  [physics.ins-det]} \BibitemShut {NoStop}%
\bibitem [{\citenamefont {Blondel}\ \emph {et~al.}(2018)\citenamefont {Blondel}
  \emph {et~al.}}]{Blondel:2017orl}%
  \BibitemOpen
  \bibfield  {author} {\bibinfo {author} {\bibfnamefont {A.}~\bibnamefont
  {Blondel}} \emph {et~al.},\ }\href {\doibase 10.1088/1748-0221/13/02/P02006}
  {\bibfield  {journal} {\bibinfo  {journal} {JINST}\ }\textbf {\bibinfo
  {volume} {13}},\ \bibinfo {pages} {P02006} (\bibinfo {year} {2018})},\
  \Eprint {http://arxiv.org/abs/1707.01785} {arXiv:1707.01785
  [physics.ins-det]} \BibitemShut {NoStop}%
\bibitem [{\citenamefont {Korzenev}\ \emph {et~al.}(2019)\citenamefont
  {Korzenev} \emph {et~al.}}]{Korzenev:2019kud}%
  \BibitemOpen
  \bibfield  {author} {\bibinfo {author} {\bibfnamefont {A.}~\bibnamefont
  {Korzenev}} \emph {et~al.},\ }\href {\doibase 10.7566/JPSCP.27.011005}
  {\bibfield  {journal} {\bibinfo  {journal} {JPS Conf. Proc.}\ }\textbf
  {\bibinfo {volume} {27}},\ \bibinfo {pages} {011005} (\bibinfo {year}
  {2019})},\ \Eprint {http://arxiv.org/abs/1901.07785} {arXiv:1901.07785
  [physics.ins-det]} \BibitemShut {NoStop}%
\bibitem [{\citenamefont {Abe}\ \emph {et~al.}(2020)\citenamefont {Abe} \emph
  {et~al.}}]{T2K:2019bcf}%
  \BibitemOpen
  \bibfield  {author} {\bibinfo {author} {\bibfnamefont {K.}~\bibnamefont
  {Abe}} \emph {et~al.} (\bibinfo {collaboration} {T2K}),\ }\href {\doibase
  10.1038/s41586-020-2177-0} {\bibfield  {journal} {\bibinfo  {journal}
  {Nature}\ }\textbf {\bibinfo {volume} {580}},\ \bibinfo {pages} {339}
  (\bibinfo {year} {2020})},\ \bibinfo {note} {[Erratum: Nature 583, E16
  (2020)]},\ \Eprint {http://arxiv.org/abs/1910.03887} {arXiv:1910.03887
  [hep-ex]} \BibitemShut {NoStop}%
\bibitem [{\citenamefont {Abe}\ \emph {et~al.}(2021)\citenamefont {Abe} \emph
  {et~al.}}]{T2K:2021xwb}%
  \BibitemOpen
  \bibfield  {author} {\bibinfo {author} {\bibfnamefont {K.}~\bibnamefont
  {Abe}} \emph {et~al.} (\bibinfo {collaboration} {T2K}),\ }\href {\doibase
  10.1103/PhysRevD.103.112008} {\bibfield  {journal} {\bibinfo  {journal}
  {Phys. Rev. D}\ }\textbf {\bibinfo {volume} {103}},\ \bibinfo {pages}
  {112008} (\bibinfo {year} {2021})},\ \Eprint
  {http://arxiv.org/abs/2101.03779} {arXiv:2101.03779 [hep-ex]} \BibitemShut
  {NoStop}%
\bibitem [{\citenamefont {Munteanu}\ \emph {et~al.}(2020)\citenamefont
  {Munteanu}, \citenamefont {Suvorov}, \citenamefont {Dolan}, \citenamefont
  {Sgalaberna}, \citenamefont {Bolognesi}, \citenamefont {Manly}, \citenamefont
  {Yang}, \citenamefont {Giganti}, \citenamefont {Iwamoto},\ and\ \citenamefont
  {Jes\'us-Valls}}]{Munteanu:2019llq}%
  \BibitemOpen
  \bibfield  {author} {\bibinfo {author} {\bibfnamefont {L.}~\bibnamefont
  {Munteanu}}, \bibinfo {author} {\bibfnamefont {S.}~\bibnamefont {Suvorov}},
  \bibinfo {author} {\bibfnamefont {S.}~\bibnamefont {Dolan}}, \bibinfo
  {author} {\bibfnamefont {D.}~\bibnamefont {Sgalaberna}}, \bibinfo {author}
  {\bibfnamefont {S.}~\bibnamefont {Bolognesi}}, \bibinfo {author}
  {\bibfnamefont {S.}~\bibnamefont {Manly}}, \bibinfo {author} {\bibfnamefont
  {G.}~\bibnamefont {Yang}}, \bibinfo {author} {\bibfnamefont {C.}~\bibnamefont
  {Giganti}}, \bibinfo {author} {\bibfnamefont {K.}~\bibnamefont {Iwamoto}}, \
  and\ \bibinfo {author} {\bibfnamefont {C.}~\bibnamefont {Jes\'us-Valls}},\
  }\href {\doibase 10.1103/PhysRevD.101.092003} {\bibfield  {journal} {\bibinfo
   {journal} {Phys. Rev. D}\ }\textbf {\bibinfo {volume} {101}},\ \bibinfo
  {pages} {092003} (\bibinfo {year} {2020})},\ \Eprint
  {http://arxiv.org/abs/1912.01511} {arXiv:1912.01511 [physics.ins-det]}
  \BibitemShut {NoStop}%
\bibitem [{\citenamefont {Abe}\ \emph {et~al.}(2011{\natexlab{b}})\citenamefont
  {Abe} \emph {et~al.}}]{Abe:2011ks}%
  \BibitemOpen
  \bibfield  {author} {\bibinfo {author} {\bibfnamefont {K.}~\bibnamefont
  {Abe}} \emph {et~al.} (\bibinfo {collaboration} {T2K}),\ }\href {\doibase
  10.1016/j.nima.2011.06.067} {\bibfield  {journal} {\bibinfo  {journal} {Nucl.
  Instrum. Meth.}\ }\textbf {\bibinfo {volume} {A659}},\ \bibinfo {pages} {106}
  (\bibinfo {year} {2011}{\natexlab{b}})},\ \Eprint
  {http://arxiv.org/abs/1106.1238} {arXiv:1106.1238 [physics.ins-det]}
  \BibitemShut {NoStop}%
\bibitem [{\citenamefont {Blondel}\ \emph {et~al.}(2020)\citenamefont {Blondel}
  \emph {et~al.}}]{Blondel:2020hml}%
  \BibitemOpen
  \bibfield  {author} {\bibinfo {author} {\bibfnamefont {A.}~\bibnamefont
  {Blondel}} \emph {et~al.},\ }\href {\doibase 10.1088/1748-0221/15/12/P12003}
  {\bibfield  {journal} {\bibinfo  {journal} {JINST}\ }\textbf {\bibinfo
  {volume} {15}},\ \bibinfo {pages} {P12003} (\bibinfo {year} {2020})},\
  \Eprint {http://arxiv.org/abs/2008.08861} {arXiv:2008.08861
  [physics.ins-det]} \BibitemShut {NoStop}%
\bibitem [{\citenamefont {Atti\'e}\ \emph {et~al.}(2021)\citenamefont {Atti\'e}
  \emph {et~al.}}]{Attie:2021yeh}%
  \BibitemOpen
  \bibfield  {author} {\bibinfo {author} {\bibfnamefont {D.}~\bibnamefont
  {Atti\'e}} \emph {et~al.},\ }\href@noop {} {\  (\bibinfo {year} {2021})},\
  \Eprint {http://arxiv.org/abs/2106.12634} {arXiv:2106.12634
  [physics.ins-det]} \BibitemShut {NoStop}%
\bibitem [{\citenamefont {Hayato}(2009)}]{Hayato:2009zz}%
  \BibitemOpen
  \bibfield  {author} {\bibinfo {author} {\bibfnamefont {Y.}~\bibnamefont
  {Hayato}},\ }\href@noop {} {\bibfield  {journal} {\bibinfo  {journal} {Acta
  Phys. Polon.}\ }\textbf {\bibinfo {volume} {B40}},\ \bibinfo {pages} {2477}
  (\bibinfo {year} {2009})}\BibitemShut {NoStop}%
\bibitem [{\citenamefont {Abe}\ \emph {et~al.}(2013)\citenamefont {Abe} \emph
  {et~al.}}]{Abe:2012av}%
  \BibitemOpen
  \bibfield  {author} {\bibinfo {author} {\bibfnamefont {K.}~\bibnamefont
  {Abe}} \emph {et~al.} (\bibinfo {collaboration} {T2K}),\ }\href {\doibase
  10.1103/PhysRevD.87.012001, 10.1103/PhysRevD.87.019902} {\bibfield  {journal}
  {\bibinfo  {journal} {Phys. Rev.}\ }\textbf {\bibinfo {volume} {D87}},\
  \bibinfo {pages} {012001} (\bibinfo {year} {2013})},\ \bibinfo {note}
  {[Addendum: Phys. Rev.D87,no.1,019902(2013)]},\ \Eprint
  {http://arxiv.org/abs/1211.0469} {arXiv:1211.0469 [hep-ex]} \BibitemShut
  {NoStop}%
\bibitem [{t2k()}]{t2kfluxurl}%
  \BibitemOpen
  \href@noop {} {}\bibinfo {howpublished} {\url{
  http://t2k-experiment.org/wp-content/uploads/T2Kflux2016.tar}},\ \bibinfo
  {note} {accessed: 2019-08-07}\BibitemShut {NoStop}%
\bibitem [{\citenamefont {Benhar}\ \emph {et~al.}(1994)\citenamefont {Benhar},
  \citenamefont {Fabrocini}, \citenamefont {Fantoni},\ and\ \citenamefont
  {Sick}}]{Benhar:1994hw}%
  \BibitemOpen
  \bibfield  {author} {\bibinfo {author} {\bibfnamefont {O.}~\bibnamefont
  {Benhar}}, \bibinfo {author} {\bibfnamefont {A.}~\bibnamefont {Fabrocini}},
  \bibinfo {author} {\bibfnamefont {S.}~\bibnamefont {Fantoni}}, \ and\
  \bibinfo {author} {\bibfnamefont {I.}~\bibnamefont {Sick}},\ }\href {\doibase
  10.1016/0375-9474(94)90920-2} {\bibfield  {journal} {\bibinfo  {journal}
  {Nucl. Phys.}\ }\textbf {\bibinfo {volume} {A579}},\ \bibinfo {pages} {493}
  (\bibinfo {year} {1994})}\BibitemShut {NoStop}%
\bibitem [{\citenamefont {Nieves}\ \emph {et~al.}(2011)\citenamefont {Nieves},
  \citenamefont {Ruiz~Simo},\ and\ \citenamefont
  {Vicente~Vacas}}]{Nieves:2011pp}%
  \BibitemOpen
  \bibfield  {author} {\bibinfo {author} {\bibfnamefont {J.}~\bibnamefont
  {Nieves}}, \bibinfo {author} {\bibfnamefont {I.}~\bibnamefont {Ruiz~Simo}}, \
  and\ \bibinfo {author} {\bibfnamefont {M.~J.}\ \bibnamefont
  {Vicente~Vacas}},\ }\href {\doibase 10.1103/PhysRevC.83.045501} {\bibfield
  {journal} {\bibinfo  {journal} {Phys. Rev.}\ }\textbf {\bibinfo {volume}
  {C83}},\ \bibinfo {pages} {045501} (\bibinfo {year} {2011})},\ \Eprint
  {http://arxiv.org/abs/1102.2777} {arXiv:1102.2777 [hep-ph]} \BibitemShut
  {NoStop}%
\bibitem [{\citenamefont {Rein}\ and\ \citenamefont
  {Sehgal}(1981)}]{Rein:1981ys}%
  \BibitemOpen
  \bibfield  {author} {\bibinfo {author} {\bibfnamefont {D.}~\bibnamefont
  {Rein}}\ and\ \bibinfo {author} {\bibfnamefont {L.~M.}\ \bibnamefont
  {Sehgal}},\ }\href {\doibase 10.1016/0370-2693(81)90706-1} {\bibfield
  {journal} {\bibinfo  {journal} {Phys. Lett.}\ }\textbf {\bibinfo {volume}
  {104B}},\ \bibinfo {pages} {394} (\bibinfo {year} {1981})},\ \bibinfo {note}
  {[Erratum: Phys. Lett.106B,513(1981)]}\BibitemShut {NoStop}%
\bibitem [{\citenamefont {Gl{\"u}ck}\ \emph {et~al.}(1998)\citenamefont
  {Gl{\"u}ck}, \citenamefont {Reya},\ and\ \citenamefont {Vogt}}]{GRV98}%
  \BibitemOpen
  \bibfield  {author} {\bibinfo {author} {\bibfnamefont {M.}~\bibnamefont
  {Gl{\"u}ck}}, \bibinfo {author} {\bibfnamefont {E.}~\bibnamefont {Reya}}, \
  and\ \bibinfo {author} {\bibfnamefont {A.}~\bibnamefont {Vogt}},\ }\href
  {\doibase 10.1007/s100529800978} {\bibfield  {journal} {\bibinfo  {journal}
  {Eur. Phys. J. C}\ }\textbf {\bibinfo {volume} {5}},\ \bibinfo {pages} {461}
  (\bibinfo {year} {1998})}\BibitemShut {NoStop}%
\bibitem [{\citenamefont {Bodek}\ and\ \citenamefont
  {Yang}(2005)}]{Bodek:2005de}%
  \BibitemOpen
  \bibfield  {author} {\bibinfo {author} {\bibfnamefont {A.}~\bibnamefont
  {Bodek}}\ and\ \bibinfo {author} {\bibfnamefont {U.~K.}\ \bibnamefont
  {Yang}},\ }\href {\doibase 10.1063/1.2122031} {\bibfield  {journal} {\bibinfo
   {journal} {AIP Conf. Proc.}\ }\textbf {\bibinfo {volume} {792}},\ \bibinfo
  {pages} {257} (\bibinfo {year} {2005})},\ \Eprint
  {http://arxiv.org/abs/hep-ph/0508007} {arXiv:hep-ph/0508007} \BibitemShut
  {NoStop}%
\bibitem [{\citenamefont {Sjostrand}(1994)}]{SJOSTRAND199474}%
  \BibitemOpen
  \bibfield  {author} {\bibinfo {author} {\bibfnamefont {T.}~\bibnamefont
  {Sjostrand}},\ }\href {\doibase
  http://dx.doi.org/10.1016/0010-4655(94)90132-5} {\bibfield  {journal}
  {\bibinfo  {journal} {Computer Physics Communications}\ }\textbf {\bibinfo
  {volume} {82}},\ \bibinfo {pages} {74 } (\bibinfo {year} {1994})}\BibitemShut
  {NoStop}%
\bibitem [{\citenamefont {Aliaga}\ \emph {et~al.}(2020)\citenamefont {Aliaga}
  \emph {et~al.}}]{Aliaga:2020rqb}%
  \BibitemOpen
  \bibfield  {author} {\bibinfo {author} {\bibfnamefont {L.}~\bibnamefont
  {Aliaga}} \emph {et~al.} (\bibinfo {collaboration} {NuSTEC}),\ }in\
  \href@noop {} {\emph {\bibinfo {booktitle} {{NuSTEC Workshop on
  Neutrino-Nucleus Pion Production in the Resonance Region}}}}\ (\bibinfo
  {year} {2020})\ \Eprint {http://arxiv.org/abs/2011.07166} {arXiv:2011.07166
  [hep-ph]} \BibitemShut {NoStop}%
\bibitem [{\citenamefont {Pinzon~Guerra}\ \emph {et~al.}(2019)\citenamefont
  {Pinzon~Guerra} \emph {et~al.}}]{PinzonGuerra:2018rju}%
  \BibitemOpen
  \bibfield  {author} {\bibinfo {author} {\bibfnamefont {E.~S.}\ \bibnamefont
  {Pinzon~Guerra}} \emph {et~al.},\ }\href {\doibase
  10.1103/PhysRevD.99.052007} {\bibfield  {journal} {\bibinfo  {journal} {Phys.
  Rev. D}\ }\textbf {\bibinfo {volume} {99}},\ \bibinfo {pages} {052007}
  (\bibinfo {year} {2019})},\ \Eprint {http://arxiv.org/abs/1812.06912}
  {arXiv:1812.06912 [hep-ex]} \BibitemShut {NoStop}%
\bibitem [{\citenamefont {Stowell}\ \emph {et~al.}(2017)\citenamefont {Stowell}
  \emph {et~al.}}]{Stowell:2016jfr}%
  \BibitemOpen
  \bibfield  {author} {\bibinfo {author} {\bibfnamefont {P.}~\bibnamefont
  {Stowell}} \emph {et~al.},\ }\href {\doibase 10.1088/1748-0221/12/01/P01016}
  {\bibfield  {journal} {\bibinfo  {journal} {JINST}\ }\textbf {\bibinfo
  {volume} {12}},\ \bibinfo {pages} {P01016} (\bibinfo {year} {2017})},\
  \Eprint {http://arxiv.org/abs/1612.07393} {arXiv:1612.07393 [hep-ex]}
  \BibitemShut {NoStop}%
\bibitem [{\citenamefont {Abe}\ \emph {et~al.}(2018{\natexlab{b}})\citenamefont
  {Abe} \emph {et~al.}}]{Abe:2018pwo}%
  \BibitemOpen
  \bibfield  {author} {\bibinfo {author} {\bibfnamefont {K.}~\bibnamefont
  {Abe}} \emph {et~al.} (\bibinfo {collaboration} {T2K}),\ }\href {\doibase
  10.1103/PhysRevD.98.032003} {\bibfield  {journal} {\bibinfo  {journal} {Phys.
  Rev.}\ }\textbf {\bibinfo {volume} {D98}},\ \bibinfo {pages} {032003}
  (\bibinfo {year} {2018}{\natexlab{b}})},\ \Eprint
  {http://arxiv.org/abs/1802.05078} {arXiv:1802.05078 [hep-ex]} \BibitemShut
  {NoStop}%
\bibitem [{\citenamefont {Lu}\ \emph {et~al.}(2018)\citenamefont {Lu} \emph
  {et~al.}}]{Lu:2018stk}%
  \BibitemOpen
  \bibfield  {author} {\bibinfo {author} {\bibfnamefont {X.~G.}\ \bibnamefont
  {Lu}} \emph {et~al.} (\bibinfo {collaboration} {MINERvA}),\ }\href {\doibase
  10.1103/PhysRevLett.121.022504} {\bibfield  {journal} {\bibinfo  {journal}
  {Phys. Rev. Lett.}\ }\textbf {\bibinfo {volume} {121}},\ \bibinfo {pages}
  {022504} (\bibinfo {year} {2018})},\ \Eprint
  {http://arxiv.org/abs/1805.05486} {arXiv:1805.05486 [hep-ex]} \BibitemShut
  {NoStop}%
\bibitem [{\citenamefont {Lu}\ \emph {et~al.}(2016)\citenamefont {Lu},
  \citenamefont {Pickering}, \citenamefont {Dolan}, \citenamefont {Barr},
  \citenamefont {Coplowe}, \citenamefont {Uchida}, \citenamefont {Wark},
  \citenamefont {Wascko}, \citenamefont {Weber},\ and\ \citenamefont
  {Yuan}}]{Lu:2015tcr}%
  \BibitemOpen
  \bibfield  {author} {\bibinfo {author} {\bibfnamefont {X.~G.}\ \bibnamefont
  {Lu}}, \bibinfo {author} {\bibfnamefont {L.}~\bibnamefont {Pickering}},
  \bibinfo {author} {\bibfnamefont {S.}~\bibnamefont {Dolan}}, \bibinfo
  {author} {\bibfnamefont {G.}~\bibnamefont {Barr}}, \bibinfo {author}
  {\bibfnamefont {D.}~\bibnamefont {Coplowe}}, \bibinfo {author} {\bibfnamefont
  {Y.}~\bibnamefont {Uchida}}, \bibinfo {author} {\bibfnamefont
  {D.}~\bibnamefont {Wark}}, \bibinfo {author} {\bibfnamefont {M.~O.}\
  \bibnamefont {Wascko}}, \bibinfo {author} {\bibfnamefont {A.}~\bibnamefont
  {Weber}}, \ and\ \bibinfo {author} {\bibfnamefont {T.}~\bibnamefont {Yuan}},\
  }\href {\doibase 10.1103/PhysRevC.94.015503} {\bibfield  {journal} {\bibinfo
  {journal} {Phys. Rev.}\ }\textbf {\bibinfo {volume} {C94}},\ \bibinfo {pages}
  {015503} (\bibinfo {year} {2016})},\ \Eprint
  {http://arxiv.org/abs/1512.05748} {arXiv:1512.05748 [nucl-th]} \BibitemShut
  {NoStop}%
\bibitem [{\citenamefont {Furmanski}\ and\ \citenamefont
  {Sobczyk}(2017)}]{Furmanski:2016wqo}%
  \BibitemOpen
  \bibfield  {author} {\bibinfo {author} {\bibfnamefont {A.~P.}\ \bibnamefont
  {Furmanski}}\ and\ \bibinfo {author} {\bibfnamefont {J.~T.}\ \bibnamefont
  {Sobczyk}},\ }\href {\doibase 10.1103/PhysRevC.95.065501} {\bibfield
  {journal} {\bibinfo  {journal} {Phys. Rev.}\ }\textbf {\bibinfo {volume}
  {C95}},\ \bibinfo {pages} {065501} (\bibinfo {year} {2017})},\ \Eprint
  {http://arxiv.org/abs/1609.03530} {arXiv:1609.03530 [hep-ex]} \BibitemShut
  {NoStop}%
\bibitem [{\citenamefont {Dolan}(2018)}]{Dolan:2018zye}%
  \BibitemOpen
  \bibfield  {author} {\bibinfo {author} {\bibfnamefont {S.}~\bibnamefont
  {Dolan}},\ }\href@noop {} {\  (\bibinfo {year} {2018})},\ \Eprint
  {http://arxiv.org/abs/1810.06043} {arXiv:1810.06043 [hep-ex]} \BibitemShut
  {NoStop}%
\bibitem [{\citenamefont {Barlow}\ and\ \citenamefont
  {Beeston}(1993)}]{Barlow:1993dm}%
  \BibitemOpen
  \bibfield  {author} {\bibinfo {author} {\bibfnamefont {R.~J.}\ \bibnamefont
  {Barlow}}\ and\ \bibinfo {author} {\bibfnamefont {C.}~\bibnamefont
  {Beeston}},\ }\href {\doibase 10.1016/0010-4655(93)90005-W} {\bibfield
  {journal} {\bibinfo  {journal} {Comput. Phys. Commun.}\ }\textbf {\bibinfo
  {volume} {77}},\ \bibinfo {pages} {219} (\bibinfo {year} {1993})}\BibitemShut
  {NoStop}%
\bibitem [{\citenamefont {Niewczas}\ and\ \citenamefont
  {Sobczyk}(2019)}]{Niewczas:2019fro}%
  \BibitemOpen
  \bibfield  {author} {\bibinfo {author} {\bibfnamefont {K.}~\bibnamefont
  {Niewczas}}\ and\ \bibinfo {author} {\bibfnamefont {J.~T.}\ \bibnamefont
  {Sobczyk}},\ }\href {\doibase 10.1103/PhysRevC.100.015505} {\bibfield
  {journal} {\bibinfo  {journal} {Phys. Rev. C}\ }\textbf {\bibinfo {volume}
  {100}},\ \bibinfo {pages} {015505} (\bibinfo {year} {2019})},\ \Eprint
  {http://arxiv.org/abs/1902.05618} {arXiv:1902.05618 [hep-ex]} \BibitemShut
  {NoStop}%
\bibitem [{\citenamefont {Dytman}\ \emph {et~al.}(2021)\citenamefont {Dytman},
  \citenamefont {Hayato}, \citenamefont {Raboanary}, \citenamefont {Sobczyk},
  \citenamefont {Tena~Vidal},\ and\ \citenamefont
  {Vololoniaina}}]{Dytman:2021ohr}%
  \BibitemOpen
  \bibfield  {author} {\bibinfo {author} {\bibfnamefont {S.}~\bibnamefont
  {Dytman}}, \bibinfo {author} {\bibfnamefont {Y.}~\bibnamefont {Hayato}},
  \bibinfo {author} {\bibfnamefont {R.}~\bibnamefont {Raboanary}}, \bibinfo
  {author} {\bibfnamefont {J.}~\bibnamefont {Sobczyk}}, \bibinfo {author}
  {\bibfnamefont {J.}~\bibnamefont {Tena~Vidal}}, \ and\ \bibinfo {author}
  {\bibfnamefont {N.}~\bibnamefont {Vololoniaina}},\ }\href@noop {} {\
  (\bibinfo {year} {2021})},\ \Eprint {http://arxiv.org/abs/2103.07535}
  {arXiv:2103.07535 [hep-ph]} \BibitemShut {NoStop}%
\bibitem [{\citenamefont {Berns}\ \emph {et~al.}(2020)\citenamefont {Berns},
  \citenamefont {Boyarintsev}, \citenamefont {Hugon}, \citenamefont {Kose},
  \citenamefont {Sgalaberna}, \citenamefont {De~Roeck}, \citenamefont
  {Lebedynskiy}, \citenamefont {Sibilieva},\ and\ \citenamefont
  {Zhmurin}}]{Berns:2020ehg}%
  \BibitemOpen
  \bibfield  {author} {\bibinfo {author} {\bibfnamefont {S.}~\bibnamefont
  {Berns}}, \bibinfo {author} {\bibfnamefont {A.}~\bibnamefont {Boyarintsev}},
  \bibinfo {author} {\bibfnamefont {S.}~\bibnamefont {Hugon}}, \bibinfo
  {author} {\bibfnamefont {U.}~\bibnamefont {Kose}}, \bibinfo {author}
  {\bibfnamefont {D.}~\bibnamefont {Sgalaberna}}, \bibinfo {author}
  {\bibfnamefont {A.}~\bibnamefont {De~Roeck}}, \bibinfo {author}
  {\bibfnamefont {A.}~\bibnamefont {Lebedynskiy}}, \bibinfo {author}
  {\bibfnamefont {T.}~\bibnamefont {Sibilieva}}, \ and\ \bibinfo {author}
  {\bibfnamefont {P.}~\bibnamefont {Zhmurin}},\ }\href {\doibase
  10.1088/1748-0221/15/10/P10019} {\bibfield  {journal} {\bibinfo  {journal}
  {JINST}\ }\textbf {\bibinfo {volume} {15}},\ \bibinfo {pages} {10} (\bibinfo
  {year} {2020})},\ \Eprint {http://arxiv.org/abs/2011.09859} {arXiv:2011.09859
  [physics.ins-det]} \BibitemShut {NoStop}%
\bibitem [{\citenamefont {Boyarintsev}\ \emph {et~al.}(2021)\citenamefont
  {Boyarintsev} \emph {et~al.}}]{Boyarintsev:2021uyw}%
  \BibitemOpen
  \bibfield  {author} {\bibinfo {author} {\bibfnamefont {A.}~\bibnamefont
  {Boyarintsev}} \emph {et~al.},\ }\href@noop {} {\  (\bibinfo {year}
  {2021})},\ \Eprint {http://arxiv.org/abs/2108.11897} {arXiv:2108.11897
  [physics.ins-det]} \BibitemShut {NoStop}%
\end{thebibliography}%

\end{document}